\documentclass[fleqn,usenatbib]{mnras}

\usepackage{newtxtext,newtxmath}
\usepackage[normalem]{ulem}

\usepackage[T1]{fontenc}
\usepackage{subcaption}
\DeclareRobustCommand{\VAN}[3]{#2}
\let\VANthebibliography\thebibliography
\def\thebibliography{\DeclareRobustCommand{\VAN}[3]{##3}\VANthebibliography}

\usepackage{graphicx}	
\usepackage{amsmath}	
\usepackage{booktabs}
\usepackage{lastpage}
\usepackage{aadefs}

\title[Comets observed through HESP]{Optical spectroscopy of comets using Hanle Echelle Spectrograph (HESP)}

\author[Aravind K. et al.,] {K Aravind$^{1}$\thanks{E-mail: aravind139@gmail.com},
Kumar Venkataramani$^{2,3}$,
Shashikiran Ganesh$^{1}$,
Arun Surya$^{4}$,
\newauthor
Thirupathi Sivarani,$^{4}$
Devendra Sahu,$^{4}$
Athira Unni$^{4}$, Anil Bhardwaj$^{1}$\\
$^{1}$Physical Research Laboratory, Ahmedabad 380009, India.\\
$^{2}$Division of Physics, Mathematics, and Astronomy, California Institute of Technology, Pasadena, CA 91125, USA\\
$^{3}$IPAC, California Institute of Technology, MS 100-22, Pasadena, CA 91125, USA\\
$^{4}$Indian Institute of Astrophysics, Bengaluru 560034, India
}
\date{Accepted 2024 February 9. Received 2024 February 2029; in original form 2023 June 10}

\pubyear{2015}

\begin{document}
\label{firstpage}
\pagerange{\pageref{firstpage}--\pageref{lastpage}}
\maketitle


\begin{abstract}
Observing the vibrational/rotational lines in a comet's optical spectrum requires high-resolution spectroscopy, as they are otherwise seen as a blended feature. To achieve this, we have obtained medium and high-resolution (R ($\lambda/\Delta \lambda$) = 30000 and 60000) spectra of several comets, including C/2015 V2 (Johnson), 46P/Wirtanen, 41P/Tuttle–Giacobini–Kresák and 38P/Stephan–Oterma, using the Hanle Echelle Spectrograph (HESP) mounted on the 2-m Himalayan Chandra Telescope (HCT) in India. The spectra effectively cover the wavelength range 3700 - 10,000 \AA, allowing us to probe the various vibrational bands and band sequences to identify the rotational lines in the cometary molecular emission. We were also able to separate the cometary Oxygen lines from the telluric lines and analyse the green-to-red (G/R) forbidden oxygen [OI] ratios in a few comets. For comets C/2015 V2, 46P, and 41P, the computed G/R ratios, 0.04$\pm$0.01, 0.04$\pm$0.01, and 0.08$\pm$0.02 respectively, point to H$_2$O being a major source of Oxygen emissions. Notably, in the second fibre pointing at a location 1000 km away from the photocenter of comet 46P, the G/R ratio reduced by more than half the value observed in the first fibre,  indicating the effects of quenching within the inner coma. We also measured the NH$_2$ ortho-to-para ratio of comet 46P to be about 3.41$\pm$0.05 and derived an ammonia ratio of 1.21$\pm$0.03 corresponding to a spin temperature of $\sim$26 K. 
With these, we present the results of the study of four comets from different cometary reservoirs using medium and high-resolution optical spectroscopy, emphasising the capabilities of the instrument for future cometary studies.
\end{abstract}




\section{Introduction} \label{sec:intro}

Comets are the primordial remnants of the proto-planetary disk that formed the Solar system \citep{colli_fragments}. These icy bodies containing the pristine material from the early protosolar nebula would have probably undergone less internal and external evolution due to their small masses and long orbital periods. 
Hence, they can provide vital clues to the early history of the Solar system. Understanding the chemical composition of a comet analysed from the various molecular emission bands in different wavelength regimes is among the different aspects of a comet that needs to be studied. 
The rotational lines of a particular vibrational band in a comet's molecular emission cannot be resolved with low-resolution spectroscopy since they will be blended with each other. In the case of some molecules (C\textsubscript{2}), the bands arising with the same change in vibrational quantum number in going from one electronic state to another have wavelengths close to each other, resulting in a blended feature (Swan bands) \citep{swanband_1857}. High resolution is required to separate various bands from this blended feature. Even much higher resolution would be required to resolve the rotational structures \citep{swamy}. \cite{c2_swanband} have given a detailed study of the (C\textsubscript{2}) Swan bands using high resolution spectroscopy. \\ From the high-resolution spectroscopic observations of comets Swift-Tuttle and Brorsen-Metcalf, \cite{brown_highres} have identified 2997 lines including those of H, C\textsubscript{2}, O, NH\textsubscript{2}, CN, C\textsubscript{3}, CH, H\textsubscript{2}O\textsuperscript{+} and CH\textsuperscript{+}. High-resolution spectra of Comet C/1996 B2 (Hyakutake) and Comet  C/1995 O1 (Hale-Bopp) were obtained by \cite{G_R_hyakutake} and \cite{Zhang_2001} respectively. Similar observations have also been done for comets 153P/Ikeya–Zhang \citep{153P_highres},  21P/Giacobini–Zinner \citep{21P_highres} and C/2000 WM1 \citep{C2000WM1_highres}. \cite{21P_highres_depleted} used high-resolution spectroscopy to illustrate what it means to be a highly depleted comet and compared it  with a comet of typical composition. Also, high-resolution spectroscopy was used by \cite{Borisov_highres} to study the emissions present in the first interstellar comet, 2I/Borisov, in detail and to establish the comet's similarity with the Solar system comets. However, the high-resolution spectra of comets still contain many unidentified lines. With the help of a resolving power of 60000, \cite{cochran} have identified 12,219 lines in the wavelength range 3800 - 10,192 \AA~ for the comet 122P/de Vico using a laboratory molecular line list. This has been the most exhaustive line list until \cite{F3_highres} observed C/2020 F3 (NEOWISE) at a resolving power of 115000 to identify and catalogue 4488 cometary emission lines in the wavelength range 3830 - 6930 \AA.\\ In this work, we discuss the results from using the Hanle Echelle Spectrograph (HESP) on the Himalayan Chandra Telescope (HCT) for the high-resolution spectroscopic observation of comets. We briefly describe the various aspects regarding the instrument, observation in variable (non-sidereal) tracking mode and data reduction in section \ref{sec:obs_red}. Section \ref{sec:discuss} briefs the various emissions detected in different comets during distinct epochs and lays down a comparison among the observed band sequences and line ratios identified with the help of the available catalogues.

\section{Observation and Analysis}\label{sec:obs_red}
 We observed multiple comets during distinct epochs using the HESP instrument mounted on the HCT operated by the Indian Astronomical Observatory (IAO), Hanle at an altitude of 4500 m. The HESP instrument is a bench-mounted, dual fibre-fed spectrograph \citep{HESP}. Using an R2 echelle grating, two cross-dispersing prisms, and a 4K$\times$4K E2V CCD, it covers an effective wavelength region 3700 - 10,000 \AA~ at two modes of spectral resolution, $\lambda/\Delta \lambda$ = 30000 (medium-res) and $\lambda/\Delta \lambda$ = 60000 (high-res). A 2.7$^{\prime\prime}$ input fibre is used in the low-resolution mode, while an image slicer helps achieve the high resolution. It possesses two pinholes separated by 1.25 mm (corresponding to $\sim$ 13 arcsecs on the sky) to incorporate multiple modes of observation like star-sky, star-calib, and calib-calib. Comet observations deploy the star-sky mode where the cometary emissions will also fill the sky fibre due to the close distance of the fibres. Hence, in principle, data from both fibres can be used to analyse the emissions from the comet at two different locations in the coma. Seperate sky frames (telescope moved about 1 degree away from the photocentre) of exposures similar to comet observation are also obtained routinely. ThAr lamp is used as a calibration lamp for wavelength calibration, while Quartz blue lamp is used for flat fielding. 

 Comets being Solar system bodies, cannot be tracked in the normal observing mode (sidereal tracking). Hence, the comet was placed on the object pinhole with the help of feedback, and its position was adjusted manually for its varying motion from time to time using the direction keys in the keystone mode. We have now developed a new tracking technique currently being implemented effectively for observing comets using the HFOSC instrument on the HCT. It uses a track file consisting of the comet's altitude, azimuth and their rates at equal intervals to point the telescope to the exact position of the comet at any given time. This same method is now being applied to HESP observations so that the continuous manual correction of the comet's position would not be required.
   \begin{figure}
\centering\includegraphics[width=0.9\linewidth]{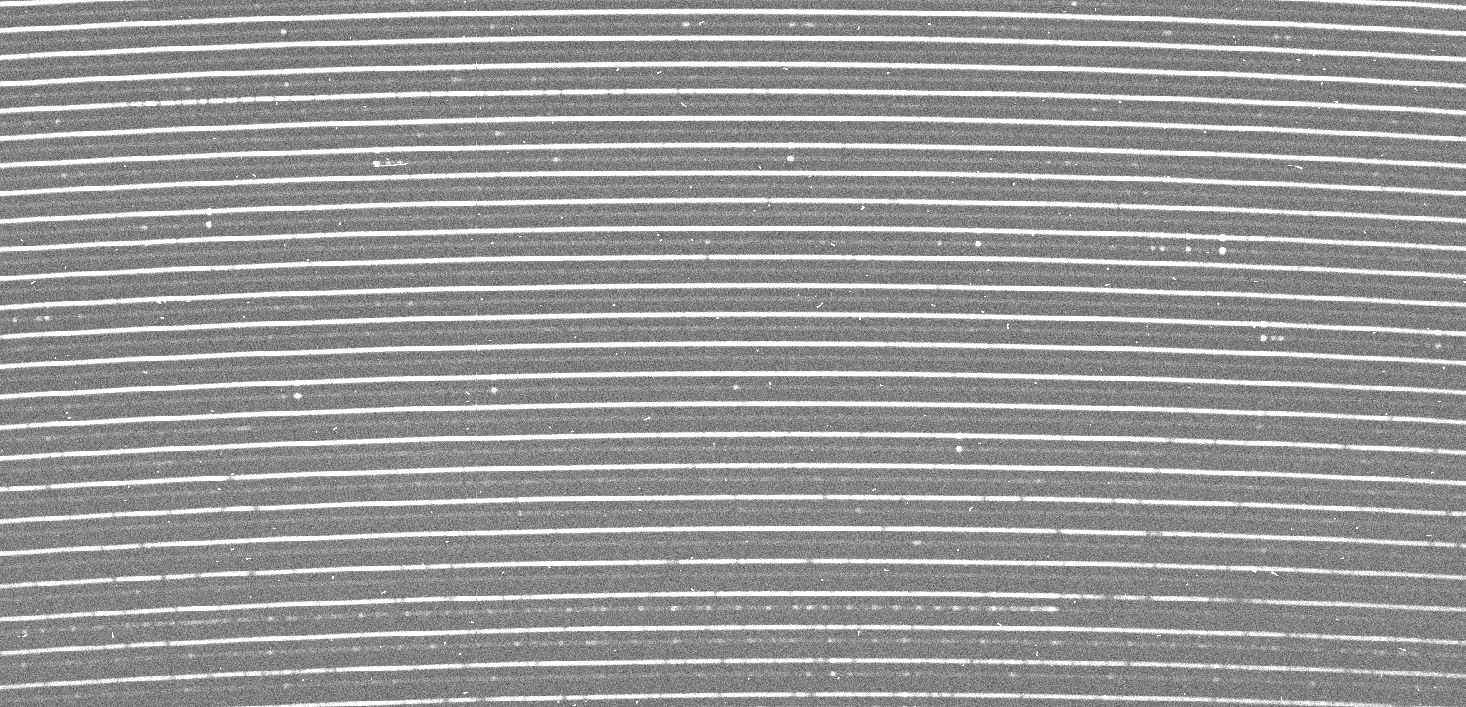}
\caption{2D image of HESP single exposure observation (raw HESP data) of Comet 46P/Wirtanen.}
\label{fig:raw_data}
\end{figure}

\begin{table}
\centering
\caption{Orbital characteristics of the observed comets and their classification}\label{tab:orbit} 
\begin{tabular}{lcccccccc}
\hline
Comet&{\textit{e}}&\textit{a}&\textit{q}&\textit{i}&\textit{P}&\textit{T$_j$}&Type\\
\hline
46P& 0.658 & 3.09 & 1.055 & 11.75 & 5.39 & 2.82 & JFC\\
C/2015 V2&  1.001 & - & 1.648 & 49.8 & - & - & LPC\\
38P & 0.8596 & 11.24 & 1.578 & 18.4 & 37.7 & 1.811 & HTC\\
41P & 0.6604 & 3.085 & 1.048 & 9.23 & 5.42 & 2.817 & JFC\\
\hline
\end{tabular}
\end{table}

\begin{table*}
\centering
\caption{Basic observational details.}\label{tab:observation} 
\begin{tabular}{clccccccc}
\hline
Object&{Date}&{Exposure time$^*$}&{$\Delta$}&$\dot{\Delta}$&{r}&$\dot{\text{r}}$&Resolution\\
&{(UT)}&(sec)&{(au)}&(km s$^{-1}$)&{(au)}&(km s$^{-1}$)&($\lambda/\Delta \lambda)$\\
\hline
46P/Wirtanen&2018-11-28T17:08:13&1800[5]&0.13&-8.32&1.07&-4.23&30000\\
&2018-12-15T20:44:25&1800[5]&0.08&-0.32&1.05&0.90&30000\\
&2018-12-28T15:56:46&1800[5]&0.11&7.6&1.08&4.70&30000\\
&2018-12-28T20:31:29 &2700[4]& & & & & 60000\\
\vspace{1mm}
&2019-01-11T21:58:57&2700[6]&0.18&10.32&1.13&8.20&30000\\
C/2015 V2 (Johnson)&2017-02-22T23:21:42&3600[5]&1.78&-21.2&2.18&-14.2&30000\\
&2017-05-02T22:33:52&1800[5]&0.99&-16.35&1.72&-7.23&30000\\
\vspace{1mm}
&2017-05-28T17:57:08&2400[4]&0.82&-4.57&1.65&-2.74&30000\\
\vspace{1mm}
38P/Stephan–Oterma&2018-11-28T23:02:27&1800[1]&0.76&-5.34&1.60&3.10&30000\\
\vspace{1mm}
41P/Tuttle–Giacobini–Kresák&2017-02-22T16:47:47&3600[2]&0.25&-10.1&1.23&-11.80&30000\\
\hline
\multicolumn{8}{l}{$^*$ Numbers in the parentheses represent the number of frames obtained for each exposure time.}\\
\hline
\end{tabular}
\end{table*}

\begin{figure*}
\centering\includegraphics[width=0.7\linewidth]{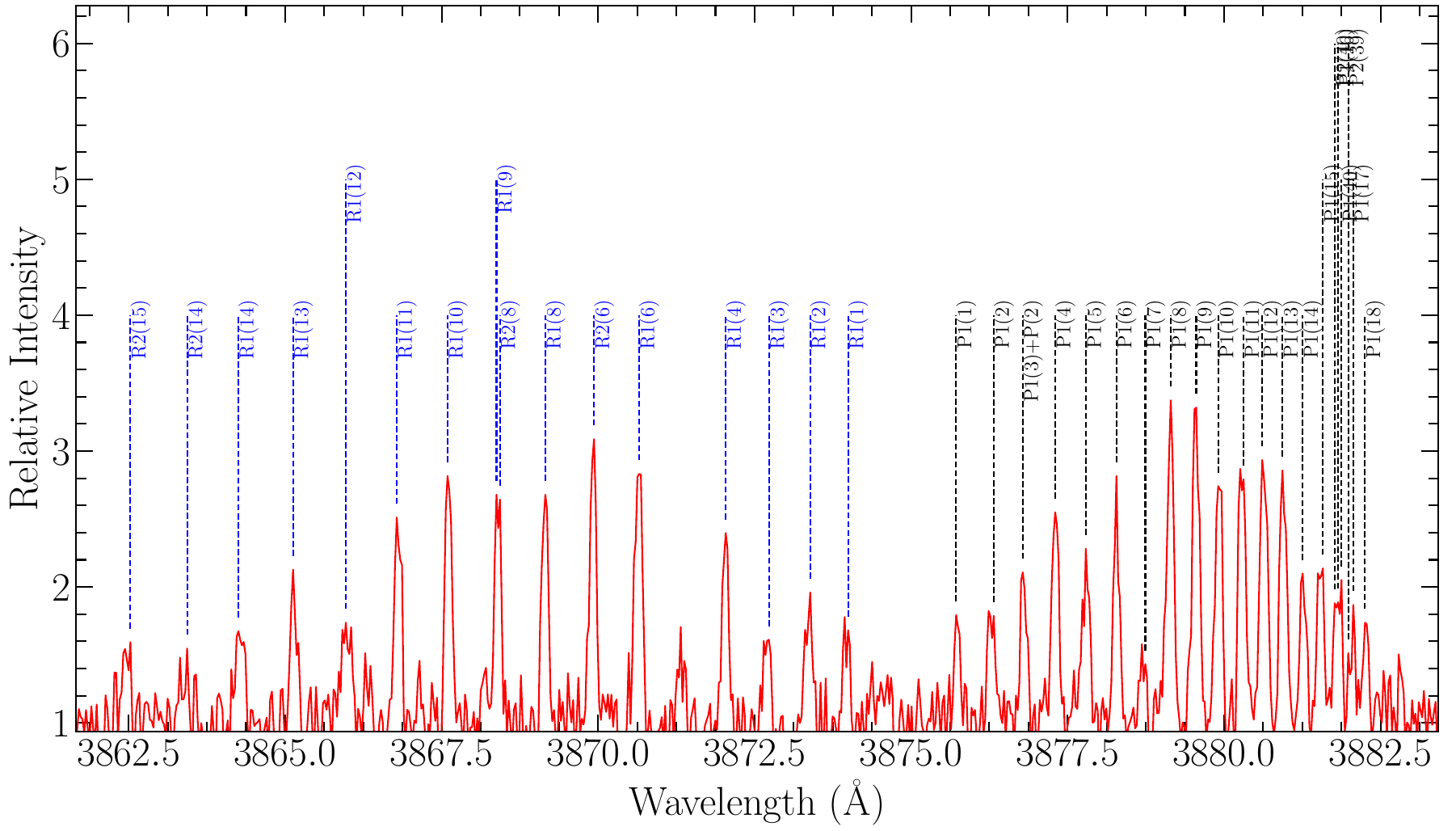}

\caption{Cross identification of the rotationa transitions present in the CN (B$^2\Sigma^+$-X$^2\Sigma^+$)(0-0) band of comet 46P observed on 2018-11-28. Blue dashed lines mark the emission lines belonging to the \textit{R} branch, while the black dashed lines mark those belonging to the \textit{P} branch.
}
\label{cn_marked}
\end{figure*}

 Most of the observations were carried out in the medium-res mode (30000). Comets being extended objects, observation in the high-res mode (60000) requires the comet to be bright enough in order to obtain the necessary Signal-to-Noise Ratio (SNR). Hence, only 46P/Wirtanen was observed in the high-resolution mode at the time of closest approach to Earth during its 2018 apparition. The complete list of observed comets along with their orbital parameters, are given in table \ref{tab:orbit} and the basic information related to the observations are provided in table \ref{tab:observation}. Multiple frames were obtained for most of the comets during each epoch. Figure \ref{fig:raw_data} illustrates a snapshot of the raw data obtained for a comet in the HESP instrument. All the observed data were reduced with the help of HESP python pipeline\footnote{\url{https://www.iiap.res.in/hesp/hesp_pipeline_manual.pdf}}. The spectrum extracted from multiple frames for each comet was combined to increase the SNR. A Doppler shift correction was incorporated with the help of \textit{dopcor} module in \textsc{Iraf} to account for the shift in emission lines due to the geocentric velocity ($\dot{\Delta}$) of the comet. Since the HESP pipeline does not perform flux calibration, each extracted order was normalised to the underlying continuum to remove the instrument response.\\

\section{Discussion}\label{sec:discuss}
In the current work, three short-period comets, 46P/Wirtanen (hereby 46P), 38P/Stephan–Oterma (hereby 38P), 41P/Tuttle–Giacobini–Kresák (hereby 41P)  and one long-period comet C/2015 V2 (Johnson) (hereby V2) were observed in the medium/high-resolution mode of HESP instrument. Emissions from various bands of CN, C$_3$, CH, C$_2$ and NH$_2$ were observed in all the comets, while emissions from CH$^+$ and CO$^+$ were not strong enough to be distinguished from the noise in the spectrum. The most prominent cometary emission features correspond to CN and C$_2$. 

Taking advantage of the very long wavelength range of the HESP instrument, we were able to observe the two different electronic band systems of CN, the violet system (B$^2\Sigma^+$-X$^2\Sigma^+$) (see section \ref{CN} and the red system (A$^2\Pi^+$-X$^2\Sigma^+$) and the Swan system of C$_2$ (see section \ref{C2}) which is most dominant in the green, orange and red region of the cometary spectrum. The lines corresponding to NH$_2$ emissions are most prominent in the red region of the spectrum but mostly blended with C$_2$ emissions when observed in low resolution ($\lambda/\delta \lambda \sim 1000$). The instrument's high resolution helps us split these emissions from the different Swan bands of C$_2$. Efforts have been made to identify the Ortho and Para lines of NH$_2$ present in the different bands. Gaussian curves are fit to the respective lines to extract the intensities and hence compute the corresponding Ortho-to-Para ratios (OPR) for certain comets (see section \ref{NH2}). The strength of the emission lines depends on the activity in the comet, which directly affects the number of lines identified (at 3$\sigma$ level).

The instrument's high resolution, aided by the comet's geocentric velocity, helps in separating the forbidden Oxygen lines of cometary origin from the corresponding telluric lines. 
Gaussian profile fitting is performed using Python (also confirmed using IRAF) to measure the intensities and FWHM of the Oxygen lines and compute the Green-to-Red doublet ratio (see section \ref{oxygen}). Additionally, a larger number of lines have been identified in certain epochs of comets 46P and C/2015 V2 owing to their brightness as compared to the fainter comets 41P and 38P (at the time of their observation). \\Even though there are infinite possibilities in using such a data set for a detailed study of the comets, this paper aims to discuss in brief the various emissions detected in different comets observed and provide a general idea regarding the capability of the HESP instrument in observing comets in high resolution.

\subsection{Line identification}
As mentioned in section \ref{sec:intro}, from the other high-resolution spectroscopic observations carried out on comets, there are few cometary emissions line catalogues available which can be used to identify the emission lines present in the comets observed in our work. Even though the molecular line list provided by \citet{F3_highres} is the latest and more exhaustive one, we have used the emission line catalogue provided by \citet{cochran} due to the similarity in the wavelength range being considered along with the comparability of the resolution used. Since certain orders of the spectrum do not have very high SNR, special care has been taken to avoid matching noise as a detected line. We have considered the standard deviation in the continuum of each order as noise (1$\sigma$) and selected those lines with signal above 3$\sigma$ level to be matched with the catalogue and identified (see figure \ref{cn_marked} for cross-identified lines). Also, similar to what has been done in \citet{F3_highres}, the catalogue lines have been matched with the observed emission line only if the wavelength coincidence was within the spectral resolution (i.e. $\pm3\Delta \lambda = \lambda/R$, where R is the resolving power of the instrument). The skylines present in the observed sky spectra have also been referred to avoid incorrect identification in regions of uncertainty.

\subsection{Detected strong emissions}\label{sec:discuss_highres}
This sub-section discusses the strong emission bands detected in the various comets.  
\subsubsection{CN(B$^2\Sigma^+$-X$^2\Sigma^+$) emission}\label{CN}
 As mentioned earlier,
 we are able to observe the violet and red systems in the CN emission. The violet band is the strongest among both detected bands and clearly shows the P and R branches of emission (see Figure \ref{cn_marked}). Swings effect \citep{swings_effect} plays a major role in the observed differences in the relative strength of the emission lines in different epochs. As mentioned in \citet{CH_CN_lines}, a few CH lines blended in the P branch of the CN emission require a much higher resolution to be resolved. The CN emission detected in the different comets observed is discussed below.\\

\begin{figure}
\centering\includegraphics[width=1\linewidth]{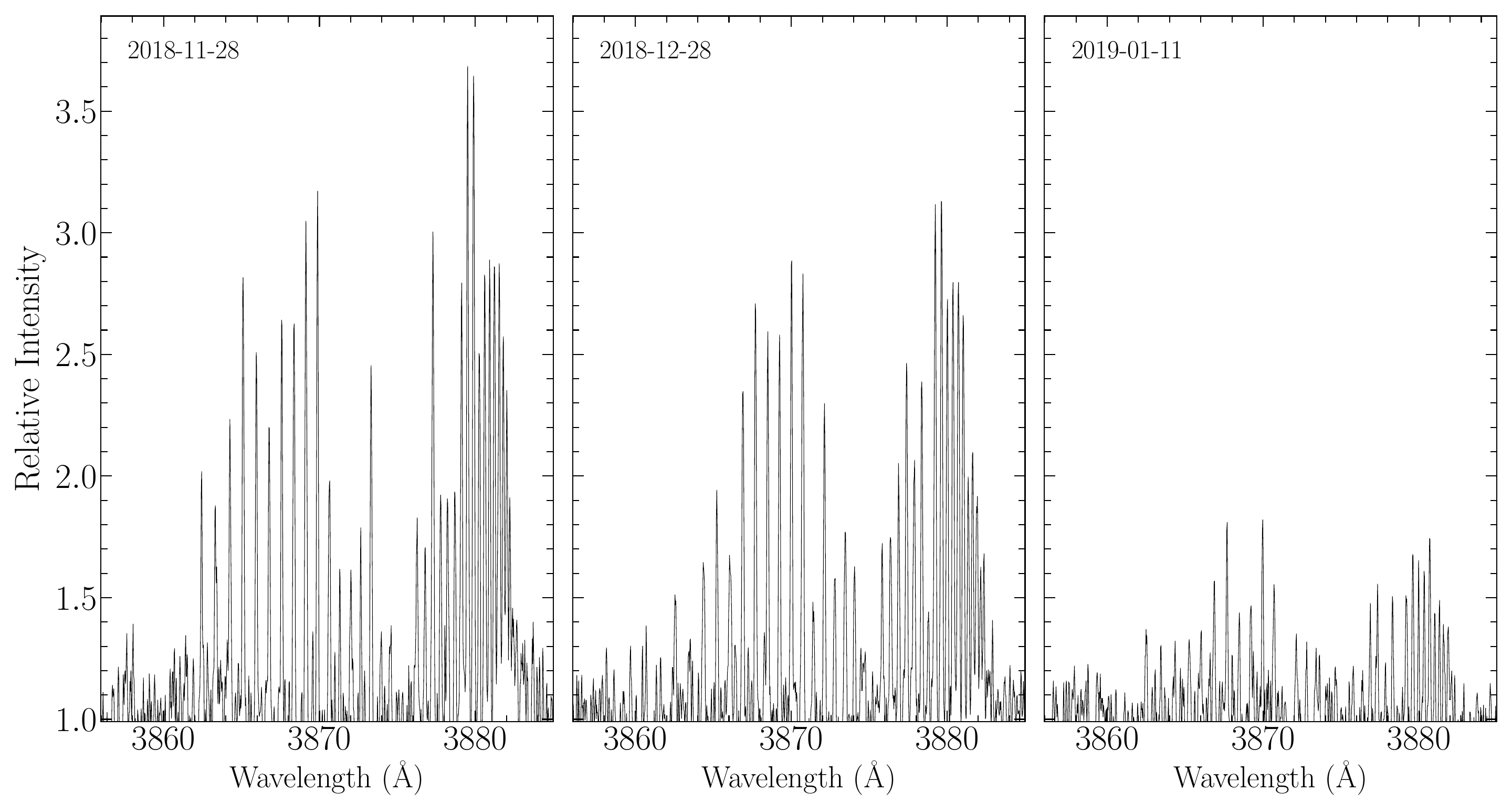}
\caption{CN($\Delta \nu=0$) emission observed in 46P during few observational epochs.}
\label{fig:46P_CN_BX_nu0}
\end{figure}

\begin{figure}
\centering\includegraphics[width=1\linewidth]{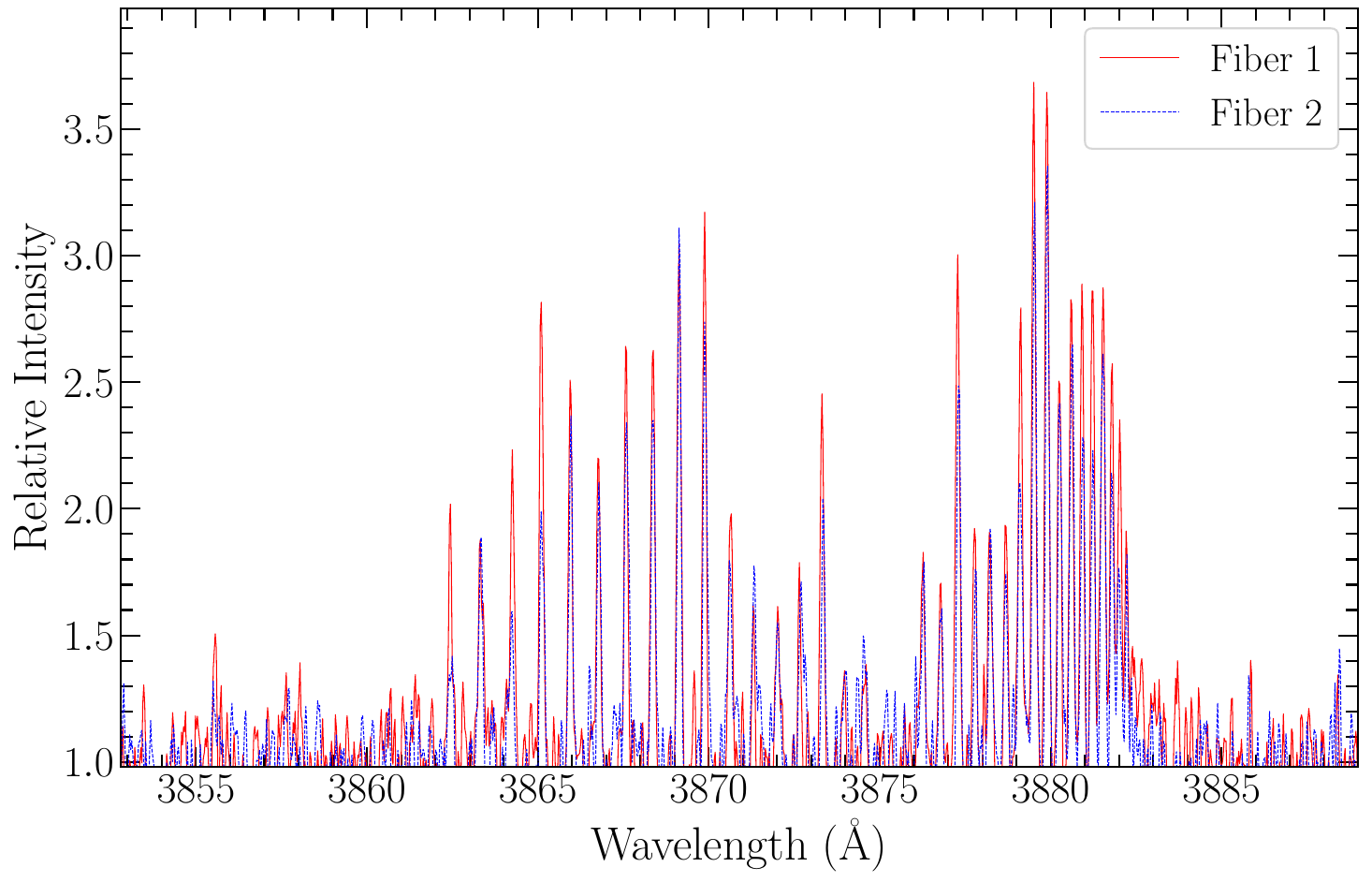}
\caption{CN($\Delta \nu=0$) band observed in Fibre 1 and Fibre 2 for 46P on 2018-11-28.}
\label{fig:46P_CN_fiber1_2}
\end{figure}

{\underline{\textbf{46P}}: }The CN B-X (0-0) emission was the strongest band observed in all the epochs (see Figure.\ref{fig:46P_CN_BX_nu0}). The observation on 2018-12-15 has been excluded in this case due to the comparatively low SNR in the band. Strong variations in the relative strengths of different emission lines are observed in spectra obtained from Fibre 1 and 2. For illustrative purpose, Figure.\ref{fig:46P_CN_fiber1_2} shows the CN band in the spectra obtained from both Fibre 1 and 2 for the epoch 2018-11-28. Such comparisons can be further used along with modelling to understand in detail the physical conditions and the photochemistry occurring at two different locations in the coma.\\

\begin{figure}
\centering\includegraphics[width=1\linewidth]{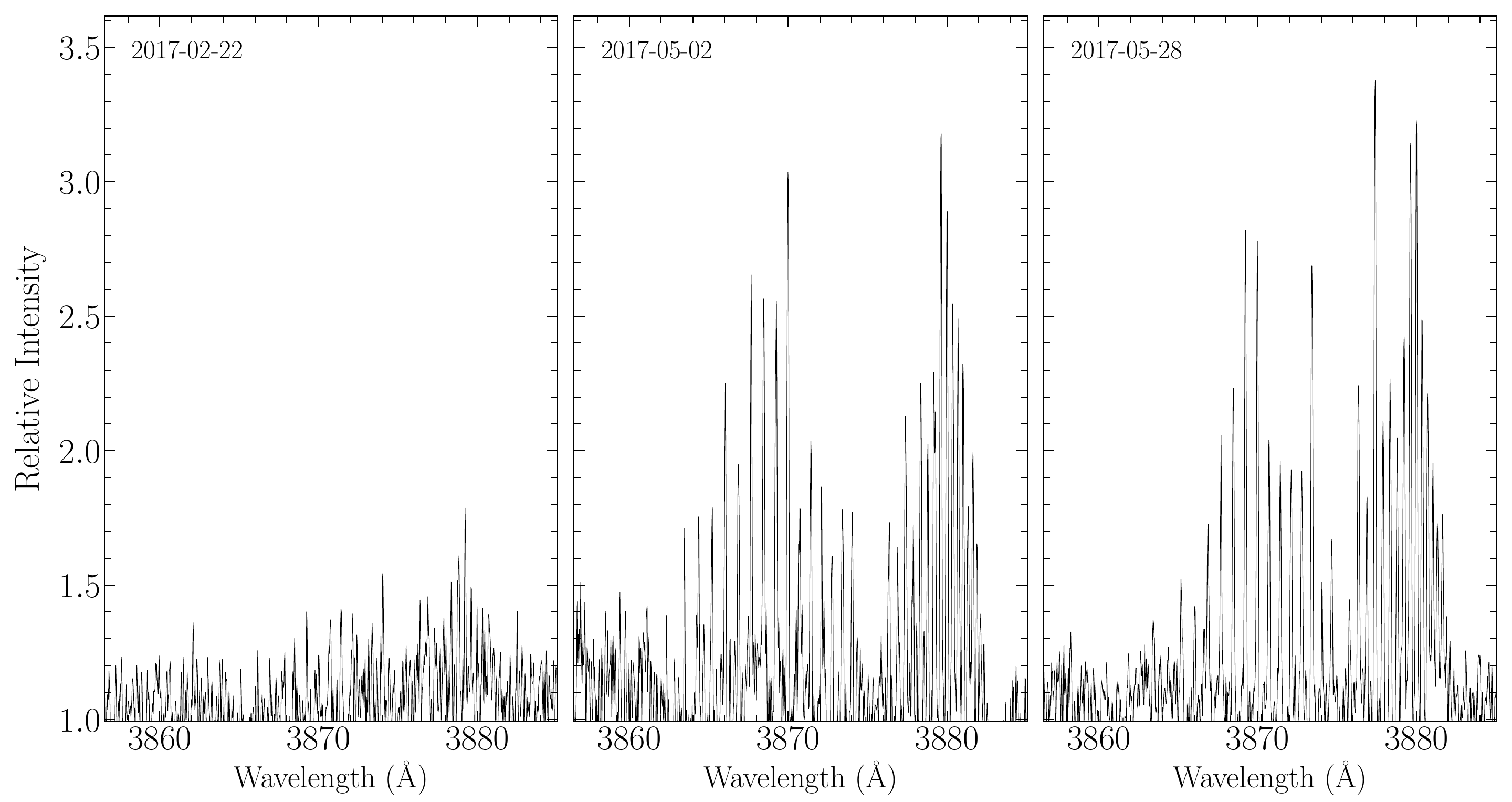}
\caption{CN($\Delta \nu=0$) emission observed in C/2015 V2 during all the observational epochs.}
\label{fig:C2015V2_CN_BX_nu0}
\end{figure}

\begin{figure}
\centering\includegraphics[width=1\linewidth]{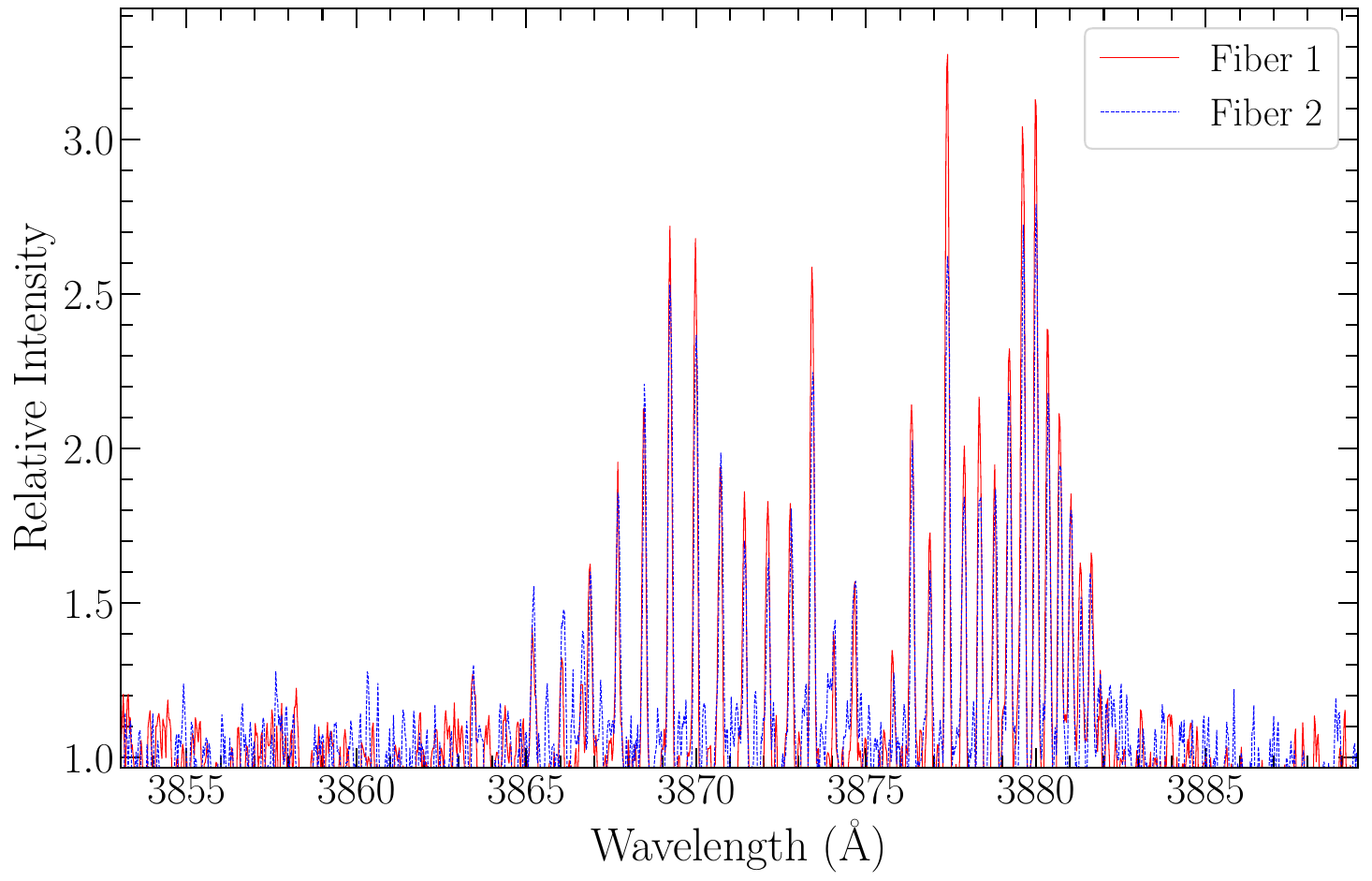}
\caption{CN($\Delta \nu=0$) band observed in Fibre 1 and Fibre 2 for C/2015 V2 on 2017-05-28.}
\label{fig:C2015V2_CN_fiber1_2}
\end{figure}

{\underline{\textbf{V2}}:} Strong emissions from the CN violet bands were observed in V2. Figure \ref{fig:C2015V2_CN_BX_nu0} represents the B-X (0-0) emission band observed across the observation dates. Even though the Swings effect plays a major role in creating differences in the relative strength of different emission lines from one epoch to another, a clear increase in the strength and definition of the P, R branches are observed as the comet approaches perihelion. Figure.\ref{fig:C2015V2_CN_fiber1_2} represents the difference in emission between spectra extracted from Fibre 1 and Fibre 2 observed on 2017-05-28. \\

{\underline{\textbf{38P and 41P}}:} Due to the comparatively weaker emission strength and the lower blue sensitivity of the instrument, the observed CN emissions were not strong enough in comets 38P and 41P as compared to 46P and V2. 

\subsubsection{C$_2$ emission}\label{C2}
For most comets observed within a heliocentric distance of 2 AU, the C$_2$ Swan band system\citep{swanband_1857, c2_swanband} is a dominant emission covering the cometary spectrum's green, orange and red regions. Most of these bands are also highly blended with various NH$_2$ bands when observed at a resolution less than $\sim$ 10,000. Various swan band systems like C$_2$($\Delta \nu =1$), C$_2$($\Delta \nu =0$), C$_2$($\Delta \nu =-1$) have been observed in most of the comets. The detection of the C$_2$($\Delta \nu =0$) Swan band region (band head 5165 \AA) observed in different comets are discussed below.\\

\begin{figure}
\centering\includegraphics[width=1\linewidth]{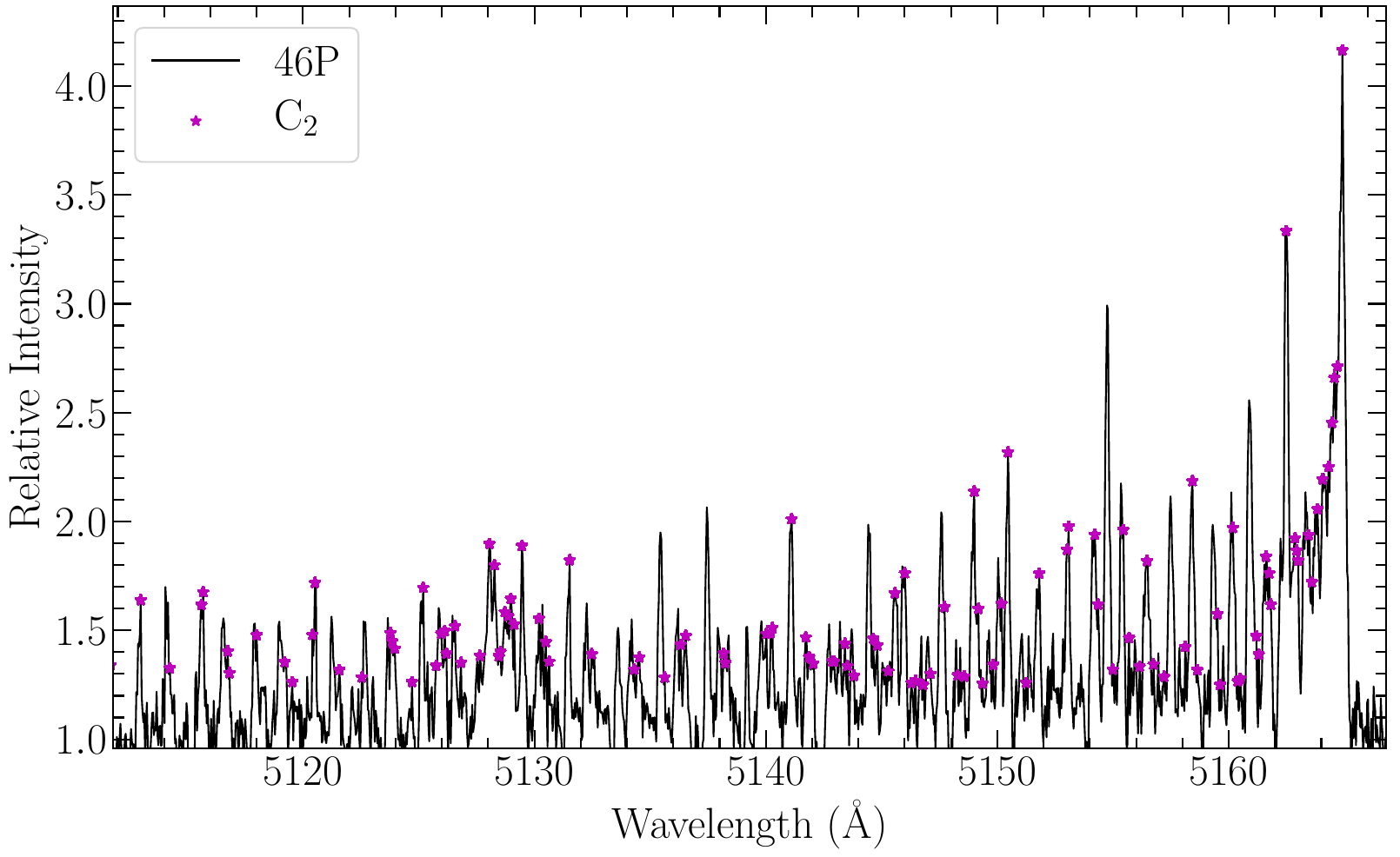}
\caption{Identified lines in the C$_2$($\Delta \nu=0$) band in 46P observed on 2018-11-28.}
\label{fig:46P_C2_identified}
\end{figure}

\begin{figure}
\centering\includegraphics[width=1\linewidth]{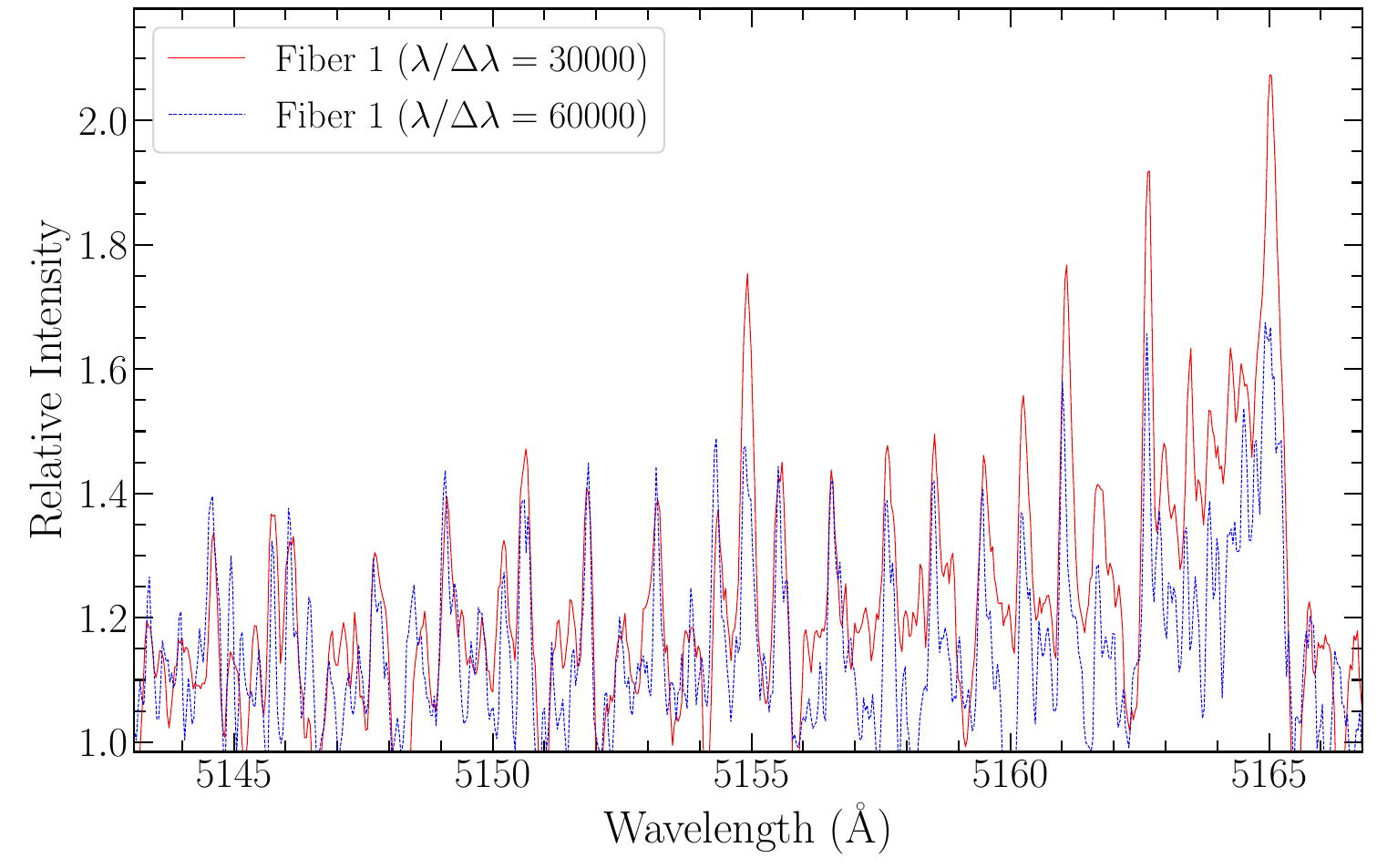}
\caption{Comparison of C$_2$($\Delta \nu=0$) band of 46P in medium resolution and high resolution.}
\label{fig:46P_C2_nu0_lowhigh}
\end{figure}

{\underline{\textbf{46P}}:} 46P was observed to be possessing strong emission strength in the C$_2$($\Delta \nu =0$) band. The C$_2$ lines identified for the observation epoch 2018-11-28 are shown in Figure \ref{fig:46P_C2_identified}. It was observed that there is NH$_2$(0-5-0) band contamination within this C$_2$ emission band. A much higher resolution would be required to deblend these emissions effectively. It is to be noted that the effect of the continuum is comparatively high in this region. Hence, proper removal of the continuum band (using the Solar spectrum) is required before comparing line ratios within the C$_2$ band between different epochs.\\
Additionally, due to the increased apparent brightness of the comet, it could be observed at a higher resolution (60000) on 2018-12-28. Certain lines observed as single lines in the medium-res mode of HESP are seen to split into multiple lines when observed in the high-res mode. Figure \ref{fig:46P_C2_nu0_lowhigh} represents one such observation in the C$_2$ emission. Only a small region of the band has been plotted for clarity. Observing a much brighter comet at this resolution could help us resolve and study the isotopic lines present in these regions.\\

\begin{figure}
\centering\includegraphics[width=1\linewidth]{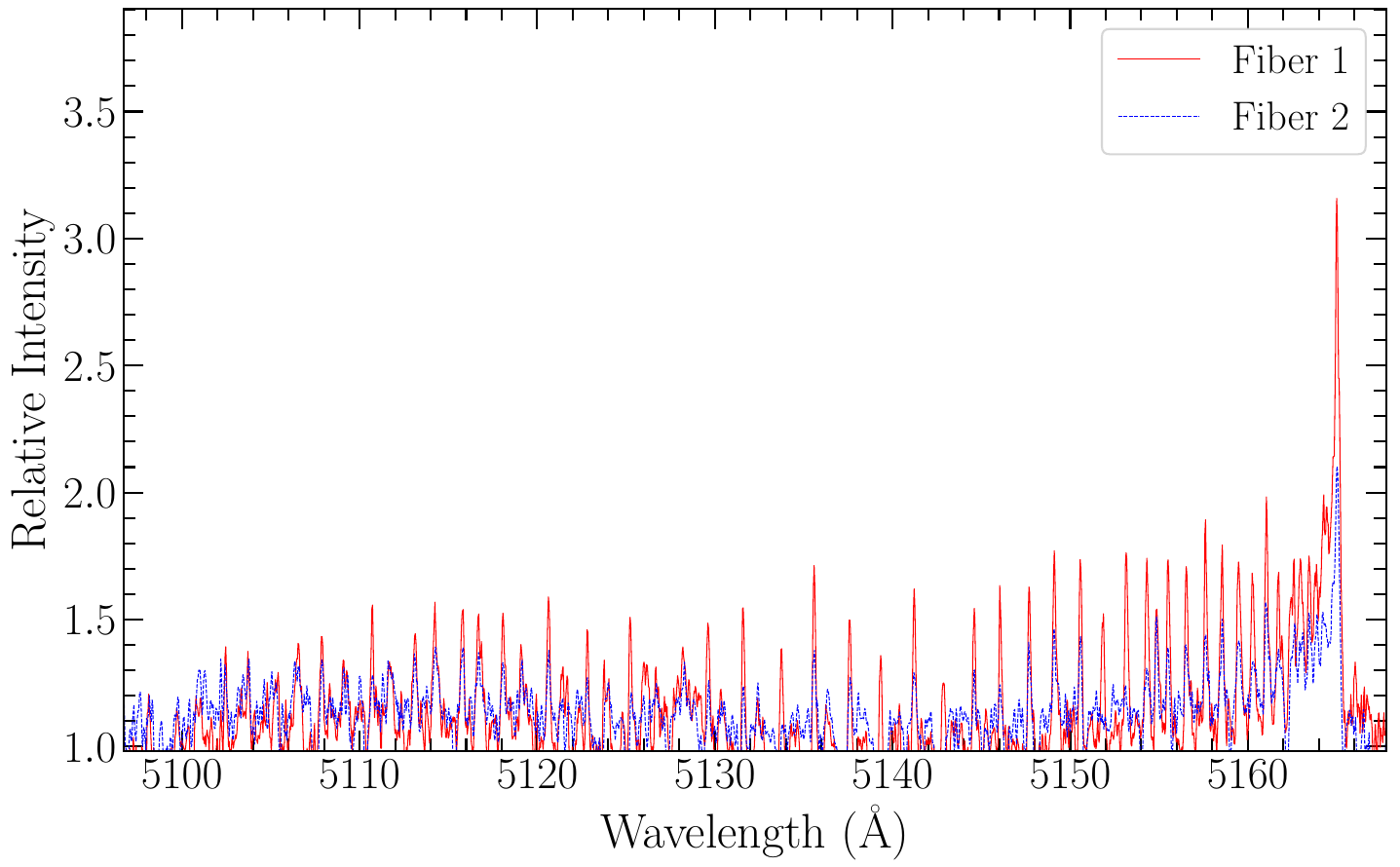}
\caption{C$_2$($\Delta \nu=0$) observed for V2 in Fibre 1 and Fibre 2 on 2018-05-28.}
\label{fig:C2015V2_C2_fiber1_2}
\end{figure}

{\underline{\textbf{V2}}:}
The comet being on its way into the inner Solar system was a major reason for the C$_2$ emission band to get stronger and more defined from one epoch to another. The major difference was seen in the band observed with the two fibres on 2018-05-28 (see Figure.\ref{fig:C2015V2_C2_fiber1_2}). Detailed studies are required to understand the observed striking difference in emission strength of various lines at the two independent locations in the coma.\\

{\underline{\textbf{38P and 41P}}:} Both the comets 38P and 41P being the fainter of the ones observed had a weaker C$_2$ band.
The intensity of the C$_2$ band head was observed to be higher in 41P than in 38P.

\subsubsection{C$_3$ emission}\label{C3}

\begin{figure}
\centering\includegraphics[width=1.01\linewidth]{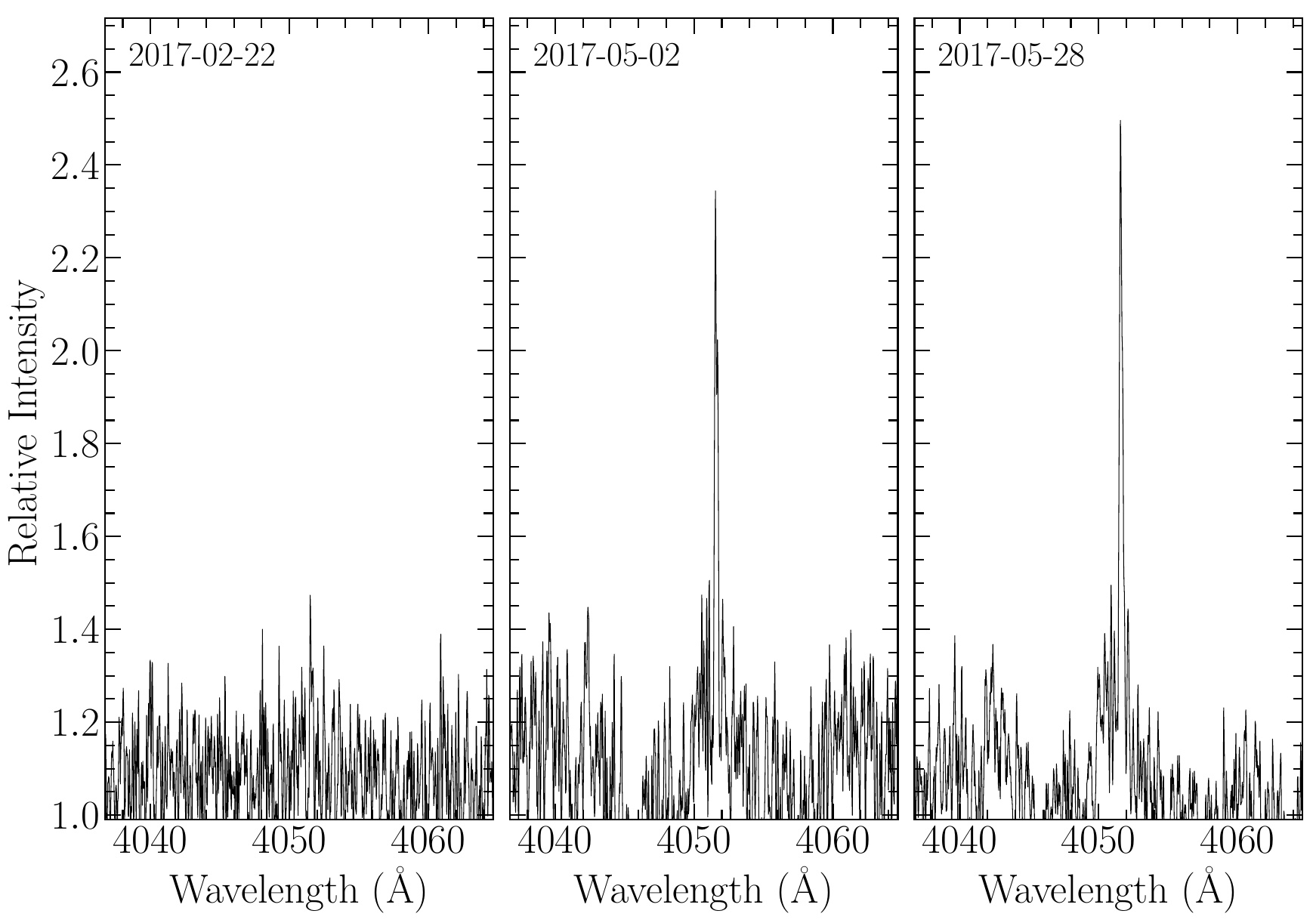}
\caption{C$_3$ emission band observed in V2 during a few epochs}
\label{fig:C2015V2_C3_all}
\end{figure}

\begin{figure}
\centering\includegraphics[width=1\linewidth]{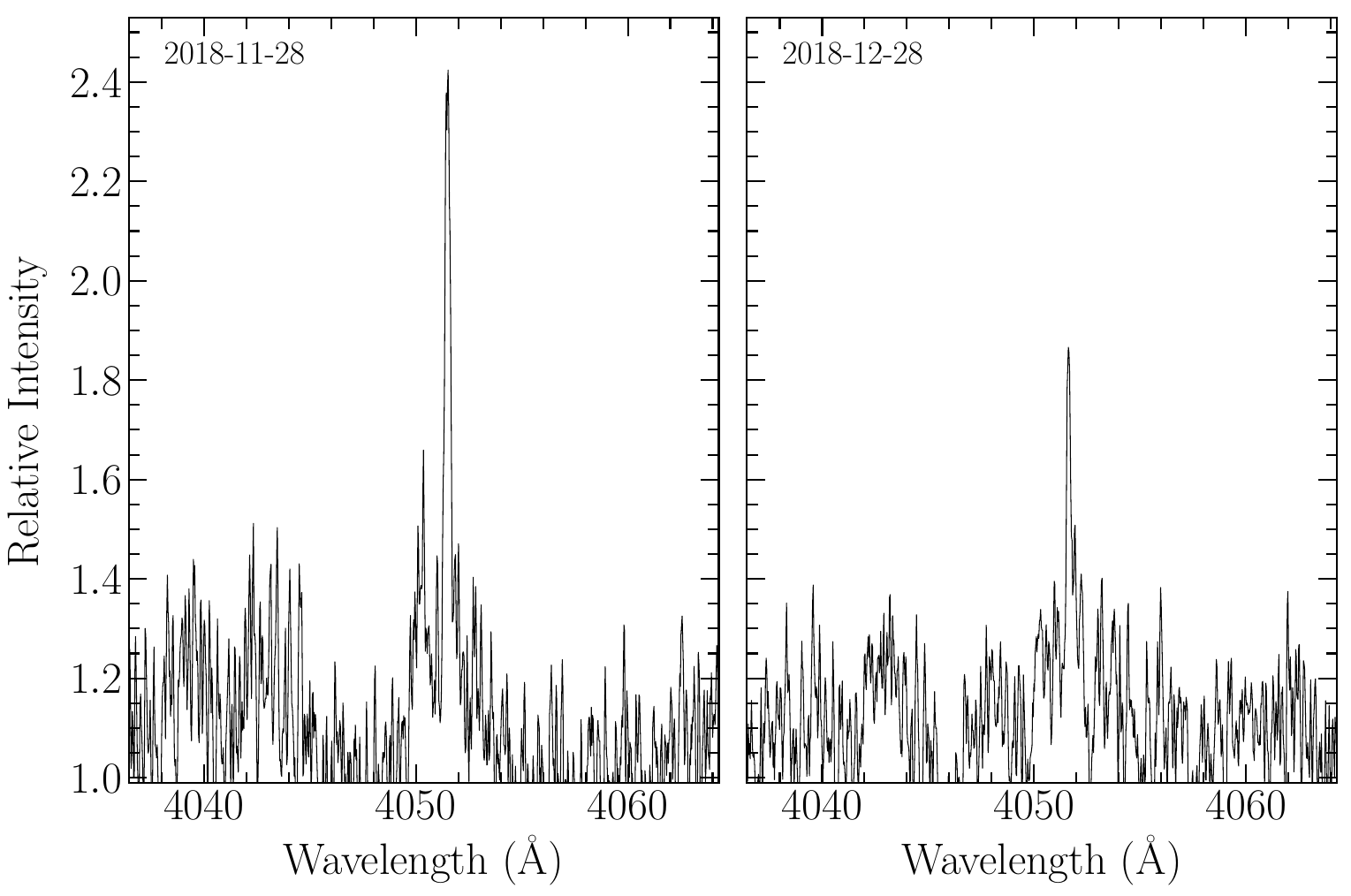}
\caption{C$_3$ emission band observed in 46P during two epochs}
\label{fig:46P_C3_al}
\end{figure}

The C$_3$ emission band was seen to be strong enough only in certain epochs of 46P and C/2015 V2. Figure.\ref{fig:C2015V2_C3_all} shows how the band's strength varies as the comet V2 approaches perihelion, while Figure.\ref{fig:46P_C3_al} illustrates the band observed in the comet 46P before and after perihelion.\\

\subsubsection{CH emission}
\begin{figure}
\centering\includegraphics[width=1\linewidth]{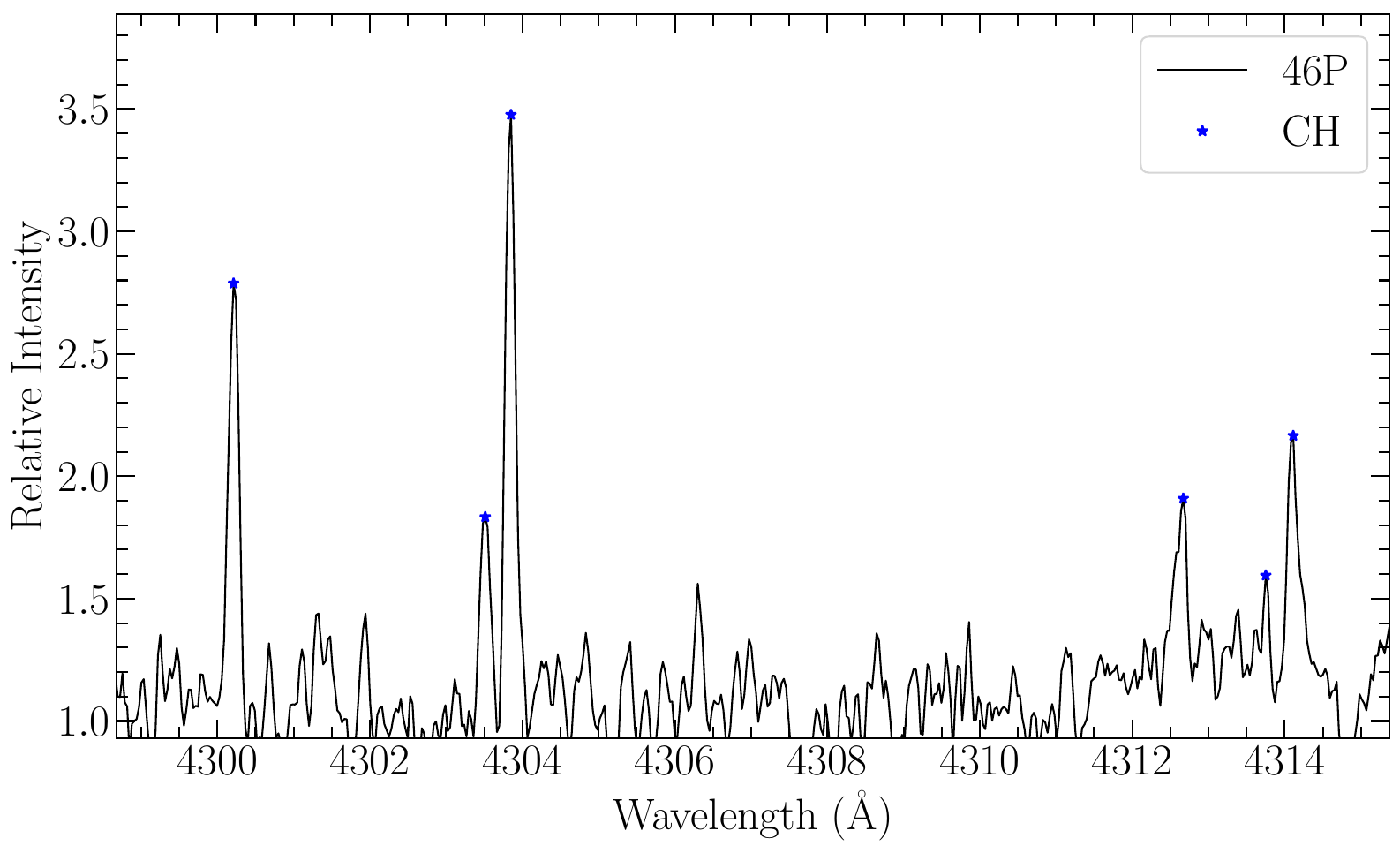}
\caption{CH emission observed in 46P on 2018-11-28}
\label{fig:46P_CH}
\end{figure}

\begin{figure}
\centering
\includegraphics[width = 1\linewidth]{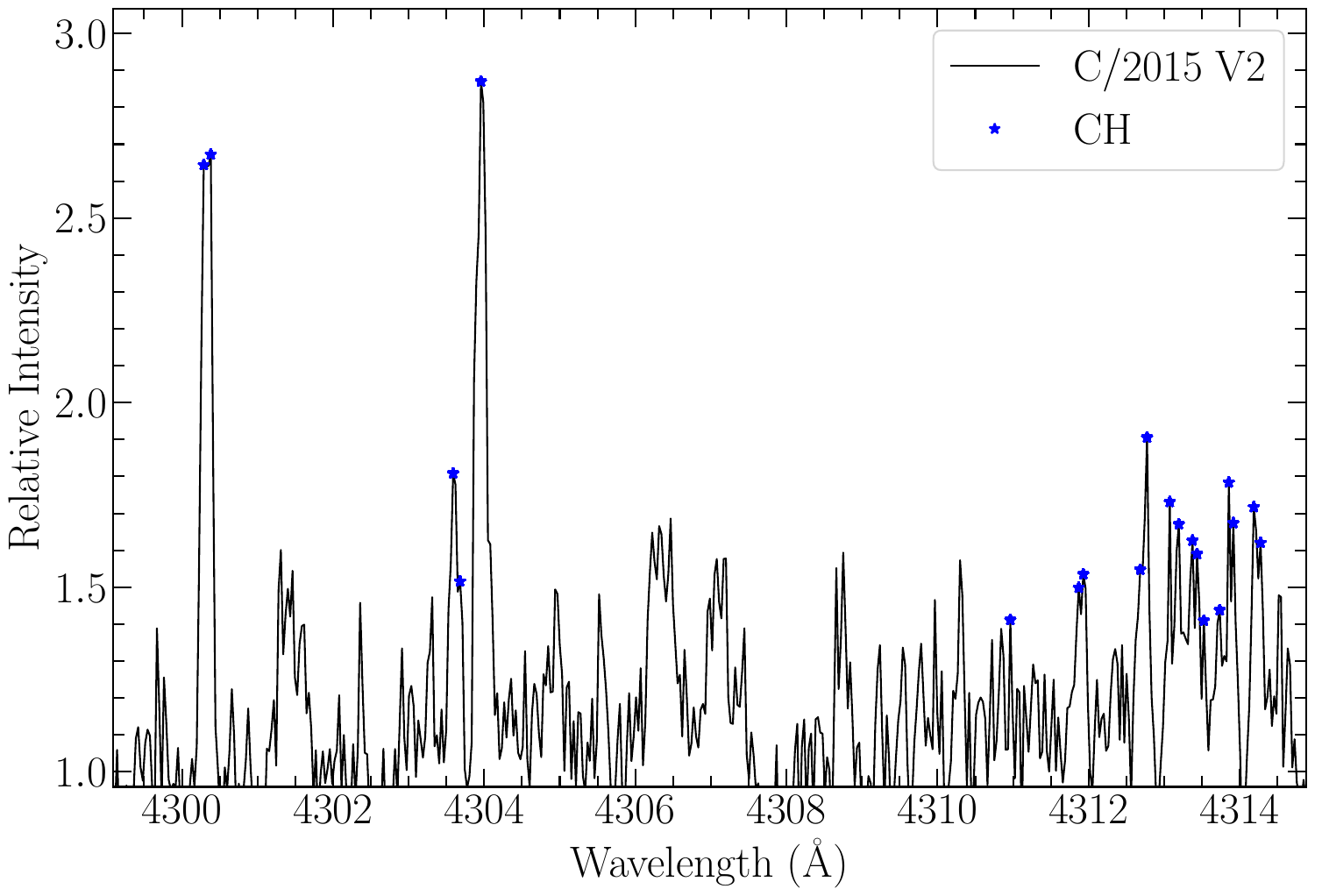}
\caption{CH observed in C/2015 V2 on 2017-05-28}
\label{fig:C2015V2_CH}
\end{figure}

CH emission observed around 4300 \AA ~is one band highly affected by the Swings effect. As a result, the line ratios vary drastically from epoch to epoch depending on the heliocentric velocity of the comet. In our case, CH emissions were clearly detected in both comets 46P and V2 (see Figures.\ref{fig:46P_CH} and \ref{fig:C2015V2_CH}). At the same time, the CH observed in V2 was seen to be strong, well-defined and similar to that observed in comet 46P, despite being at a larger heliocentric distance. Although the direct reason for such a strong detection is to be understood in detail, it could be possible that the parent molecule responsible for the CH emission is more abundant on the surface of comet V2 than in 46P , a short period comet having already made multiple passages into the inner Solar system. 

\subsubsection{NH$_2$ emissions}\label{NH2}
 NH$_2$ emissions from different transitions are observed in plenty with good SNR in all the comets. It is observed to be present along a large range of wavelength blended amongst the C$_2$ Swan bands. Since ammonia is considered the sole parent of NH$_2$, the OPR of ammonia is usually computed from the OPR of different transitions of NH$_2$ \citep{Kawakita_NH2090_spintemp_2004,21P_highres, OPR_26comets}. Even though the real implication of ammonia OPR is not yet known, the possibility of deriving the nuclear spin temperature of ammonia from OPR provides us with an opportunity to understand the conditions in the Solar nebula at the time of the formation of the cometary material or the physio-chemical conditions in the inner-most coma or beneath the surface \citep{Kawakita_NH2090_spintemp_2004,21P_highres}.\\
 
\begin{figure}
\centering
\includegraphics[width = 1\linewidth]{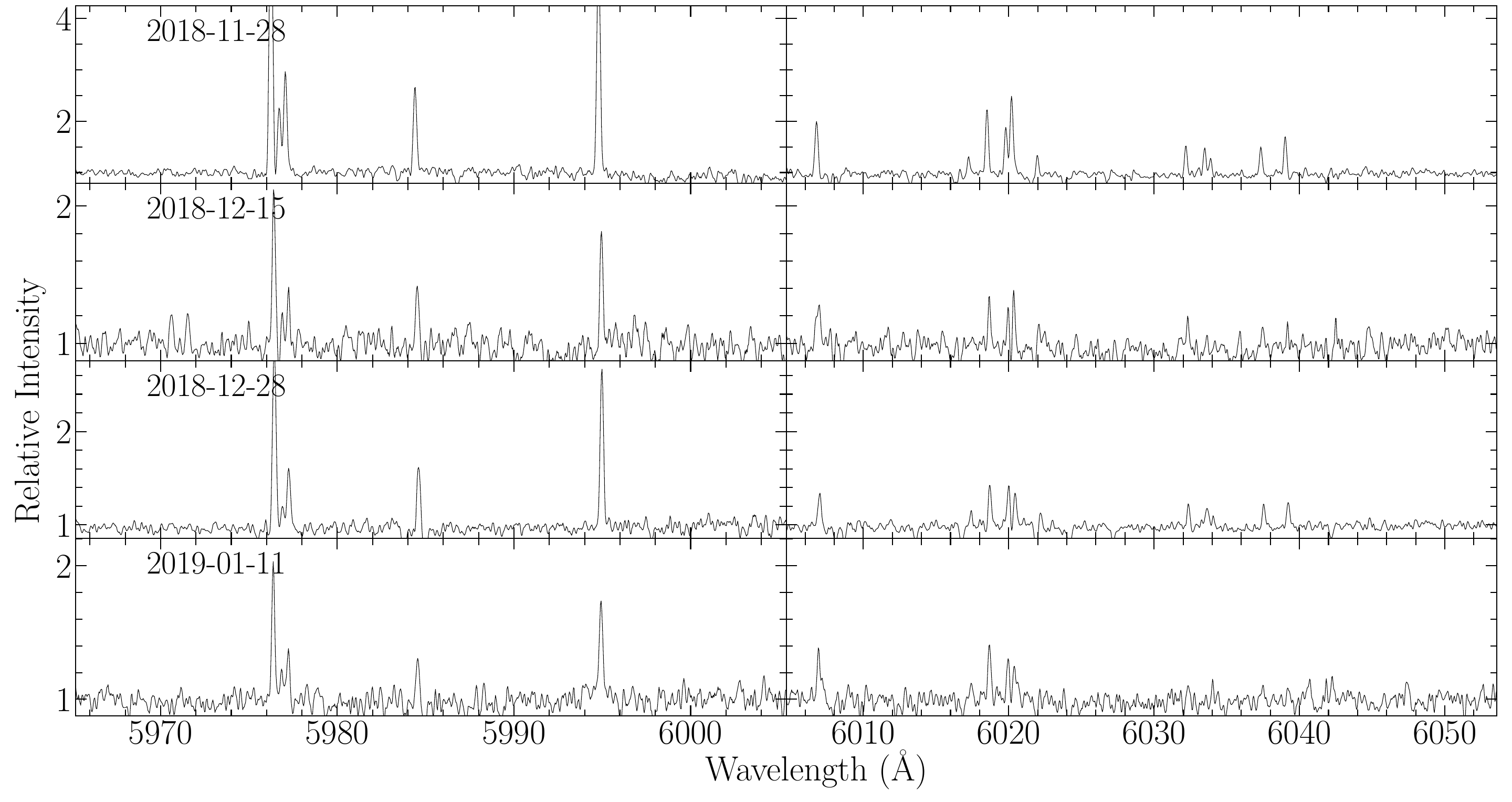}
\caption{{NH$_2$(0-9-0) emission band region in 46P on different epochs }}
\label{fig:46P_NH2_0_9_0}
\end{figure}

\begin{figure}
\centering
\includegraphics[width = 1\linewidth]{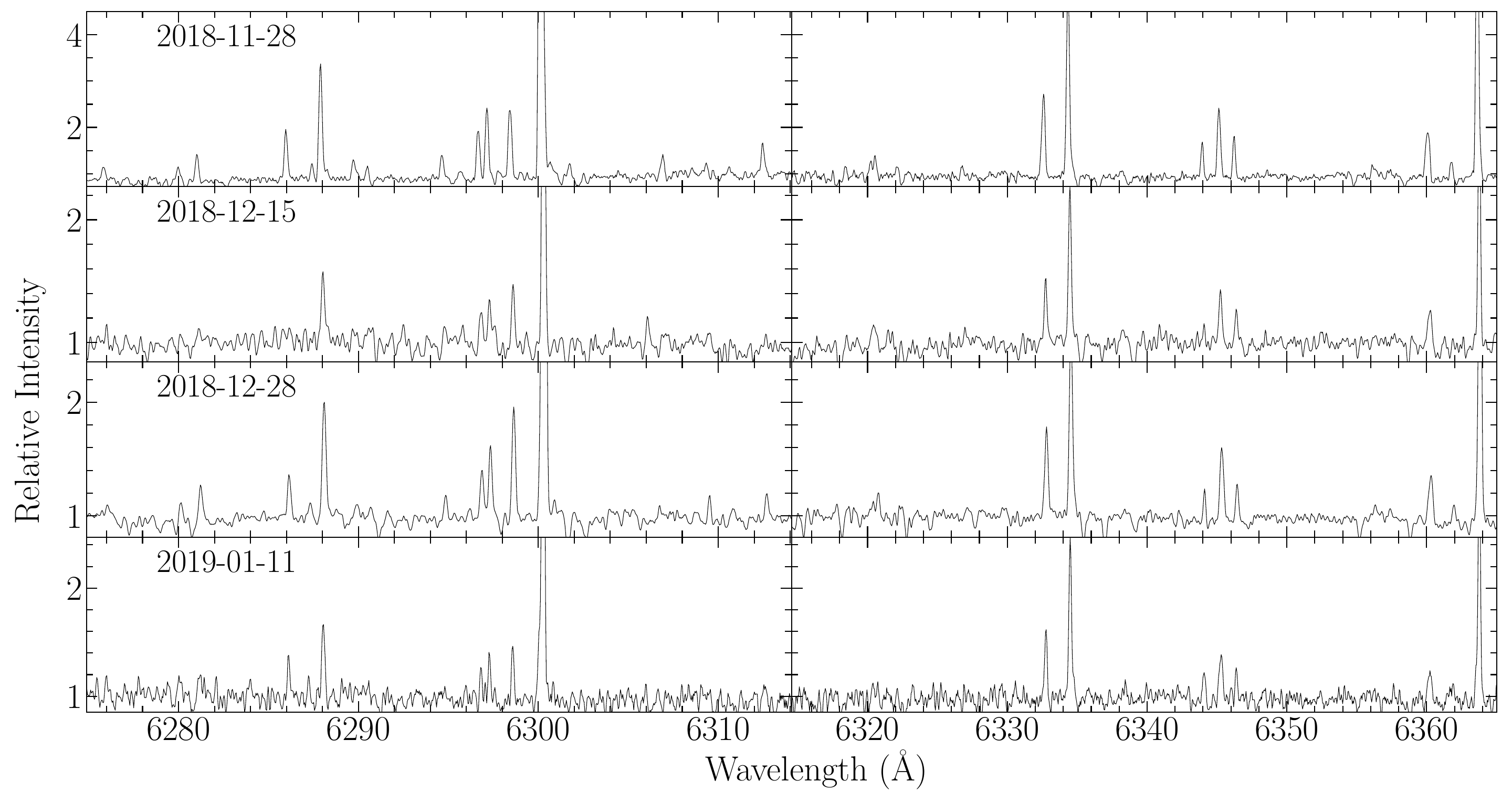}
\caption{{NH$_2$(0-8-0) emission band region in 46P on different epochs}}
\label{fig:46P_NH2_0_8_0}
\end{figure}

{\underline{\textbf{46P}}: } 
46P had the closest approach  (in about four centuries) to the Earth during this apparition, and this gave the biggest opportunity to study the various bands associated with the NH$_2$ emission present in the comet. Figures \ref{fig:46P_NH2_0_9_0} and \ref{fig:46P_NH2_0_8_0} represent the NH$_2$ (0-9-0) and (0-8-0) transitions present in the comet. With the help of the line lists given in \citet{Kawakita_NH2090_spintemp_2004} and \citet{21P_highres}, the Ortho and Para lines were identified and Gaussian curves were fit to extract the intensities of the corresponding lines to perform a basic computation of the OPR of NH$_2$ as given in Table.\ref{OPR_table}. 

  \begin{table}
\centering
\setlength\tabcolsep{10pt}
\renewcommand{\arraystretch}{1}
\caption{NH$_2$ OPR measured in 46P for different epochs. }
\setlength\tabcolsep{3pt}
\begin{tabular}{ c c c  c  }
\hline
    Date of observation & Band & Fibre 1 & Fibre 2\\ 
\hline
 2018-11-28 & (0-9-0) & 3.41 $\pm$ 0.12 & 3.48 $\pm$ 0.15\\
& (0-8-0) &  3.18 $\pm$ 0.10 & --\\
\hline
2018-12-28/High-res & (0-9-0) &  3.14 $\pm$ 0.18 & --\\
2018-12-28/medium-res &(0-9-0) & 3.84 $\pm$ 0.12 & --\\
\hline
Average & & 3.41 $\pm$ 0.05 \\
\hline
\end{tabular}
\label{OPR_table}
\end{table}

 Wherever available, the NH$_2$ bands from Fibre 2 were also used to compute the corresponding OPR. From the average observed OPR of NH$_2$, the OPR of NH$_3$ was computed to be 1.21 $\pm$ 0.03 through the relation mentioned in \cite{OPR_26comets}. This observed OPR(NH$_3$) is comparable to that reported for the same comet by \cite{46P_jehin_highres}, as well as similar to that of certain comets reported in \cite{OPR_26comets} and Hale-Bopp in \cite{Kawakita_NH2090_spintemp_2004}. This computed OPR(NH$_3$) would correspond to a nuclear spin temperature of $\sim$ 26 K. The error in spin temperature cannot be estimated in the present case since it is generalised by comparison with numbers from the literature. The exact OPR and nuclear spin temperature can only be derived with the help of precise modelling techniques. As mentioned in \cite{kawakita_9P}, further understanding of the implication of nuclear spin temperature in comets is required to have a comparative study of these values obtained in different comets. Whether a major difference in spin temperature between different dynamical classes is observed or not could vary the interpretation of these results.\\

\begin{figure}
\centering\includegraphics[width=1\linewidth]{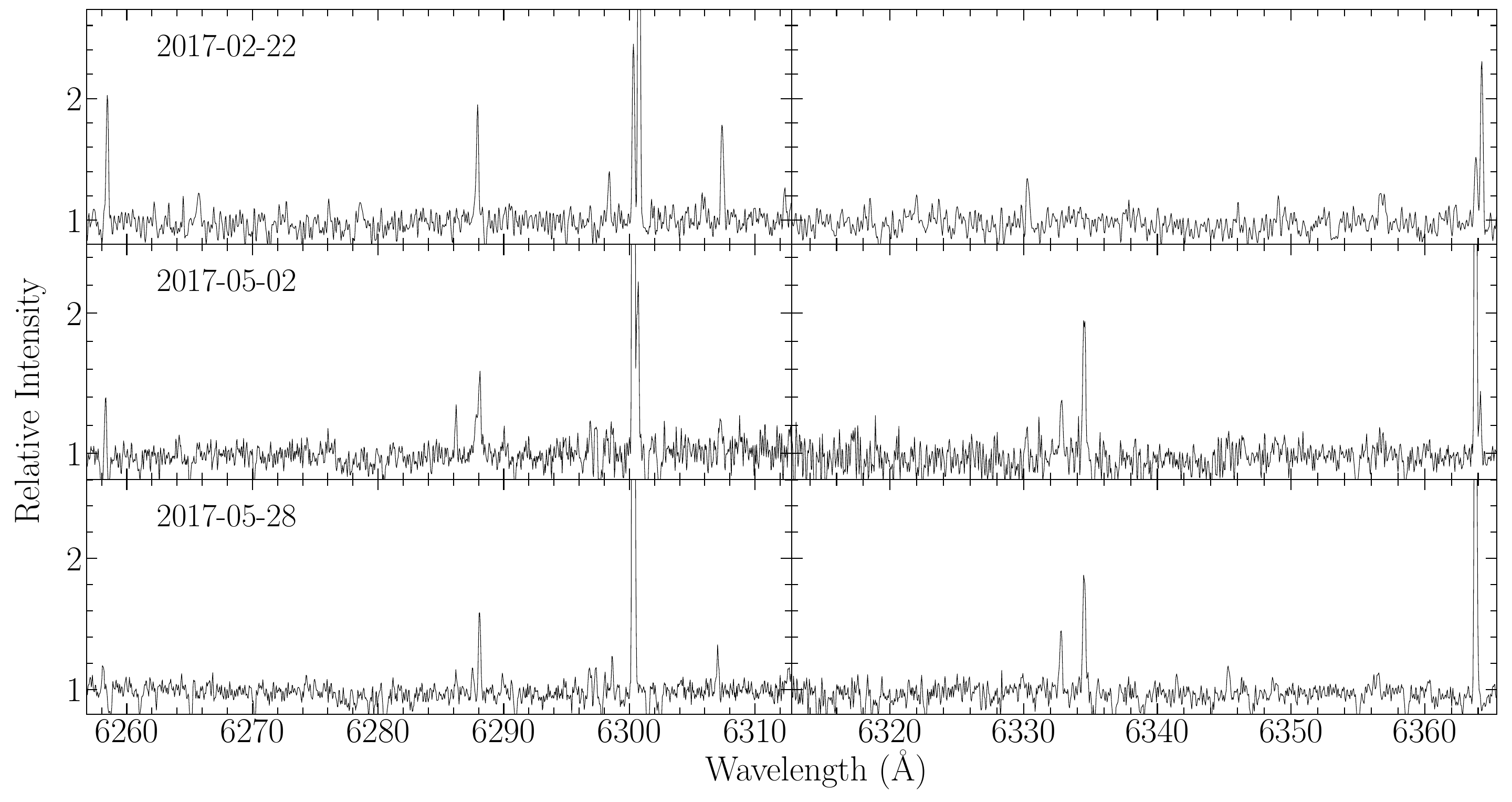}
\caption{NH$_2$(0-8-0) emission band region observed in C/2015 V2 on different epochs}
\label{fig:C2015V2_NH2_0_8_0}
\end{figure}

{\underline{\textbf{V2}}: }The NH$_2$ (0-8-0) and (0-9-0) were observed in comet V2 for all the epochs (see Figure.\ref{fig:C2015V2_NH2_0_8_0} for the observed (0-8-0) band), but the intensity in the individual Ortho and Para lines were not strong enough as in the case of 46P to perform the basic computation of OPR. Proper modelling techniques similar to those done in \cite{21P_Jehin}, \cite{21P_highres} and \cite{Kawakita_NH2090_spintemp_2004} will have to employed to synthesise the NH$_2$ spectrum matching the observed one using which the OPR and hence the spin temperature can be computed.  \\

{\underline{\textbf{38P and 41P}}: } Similar to V2, the NH$_2$ (0-8-0) and (0-9-0) were detected in comets 38P and 41P. Even though the computation of OPR would be difficult in the case of these comets, the difference in the relative strength of various NH$_2$ lines between 38P and 41P can be used to differentiate the composition of a Short Period Comet (SPC) and a Halley Type Comet (HTC). Further, modelling techniques can be incorporated to compute their respective OPR and have a comparative study.

\subsubsection{[OI]}\label{oxygen}
The most abundant materials in comets, H$_2$O, CO$_2$ and CO, all contain Oxygen, making it the most abundant element in comets. As the photodissociation of all these molecules can produce Oxygen atoms, \cite{Festour_Feldman_1981} suggested that the strength of Oxygen emission lines present at 5577.339 \AA, 6300.304 \AA~ and 6363.776 \AA~ can be used to analyse the basic compositional characteristics of a comet's nucleus.

The forbidden oxygen emission lines, the Green (G) and Red doublet (R), originate from the O($^1$S) and O($^1$D) levels, respectively. Later, work carried out by \cite{Bhardwaj_Haider_2002}, \cite{Bhardwaj_Raghuram_2012} and \cite{Raghuram_2014} defined that CO$_2$ and H$_2$O molecules contributed to the production of the green line, while only H$_2$O molecule contributed to the red doublets. Hence, a ratio known as the G/R ratio was defined, which is the ratio of the intensity of the green line to the sum of red doublets, to imply the possible major source of the detected Oxygen lines.

  \begin{figure}
\centering\includegraphics[width=1\linewidth]{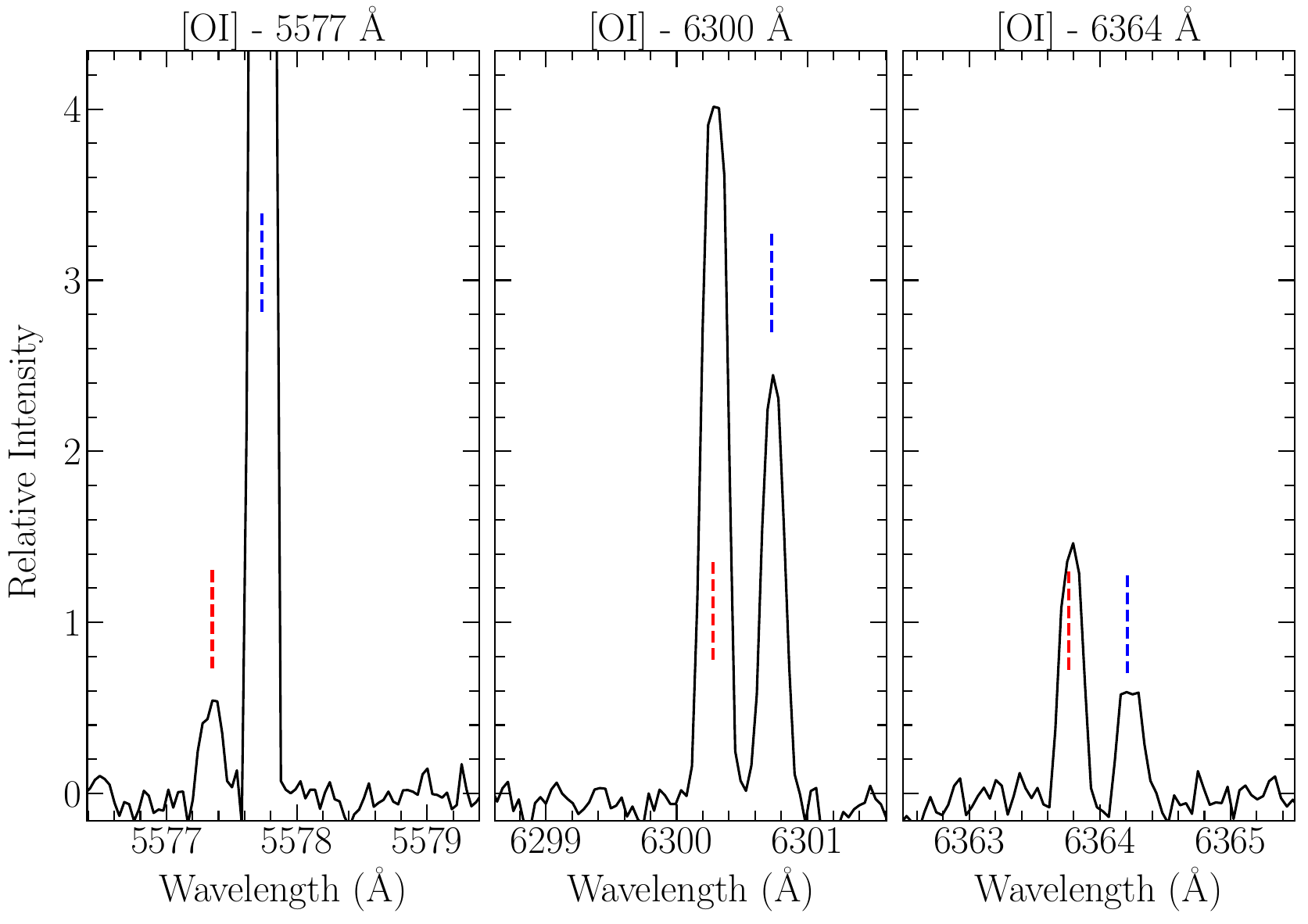}
\caption{Green and red doublet [OI] forbidden lines observed in C/2015 V2 on 2017-02-22 through the medium-resolution mode. The red and blue dashed lines mark the cometary and telluric oxygen lines, respectively.}
\label{fig:C2015V2_OI_fiber2}
\end{figure}

\begin{figure*}
\begin{subfigure}{0.33\textwidth}
\centering
\includegraphics[width = 1\linewidth]{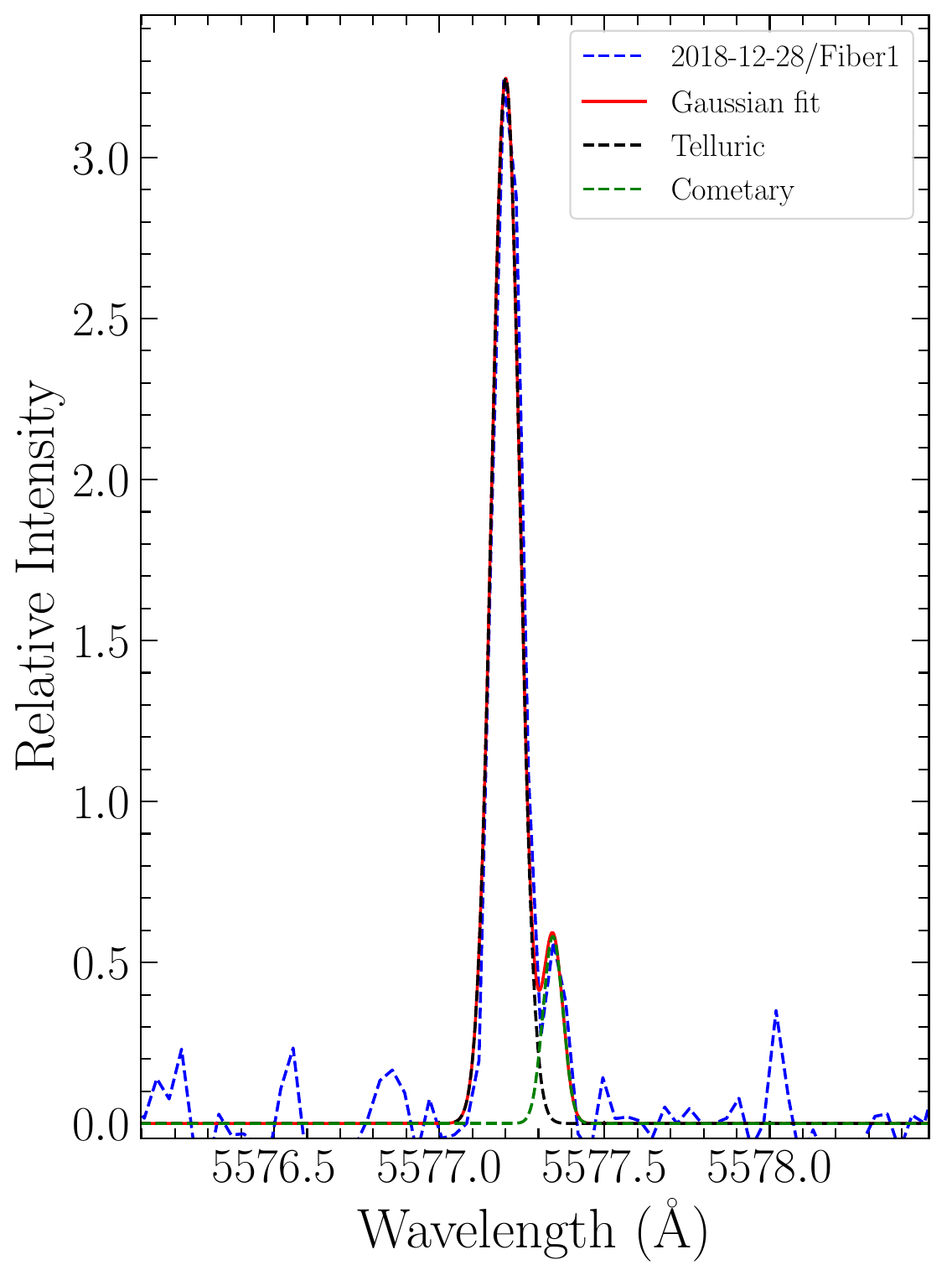}
\subcaption{{[OI] 5577 \AA}}
\label{fig:46P_O_G}
\end{subfigure}\hfill
\begin{subfigure}{0.33\textwidth}
\centering
\includegraphics[width = 0.96\linewidth]{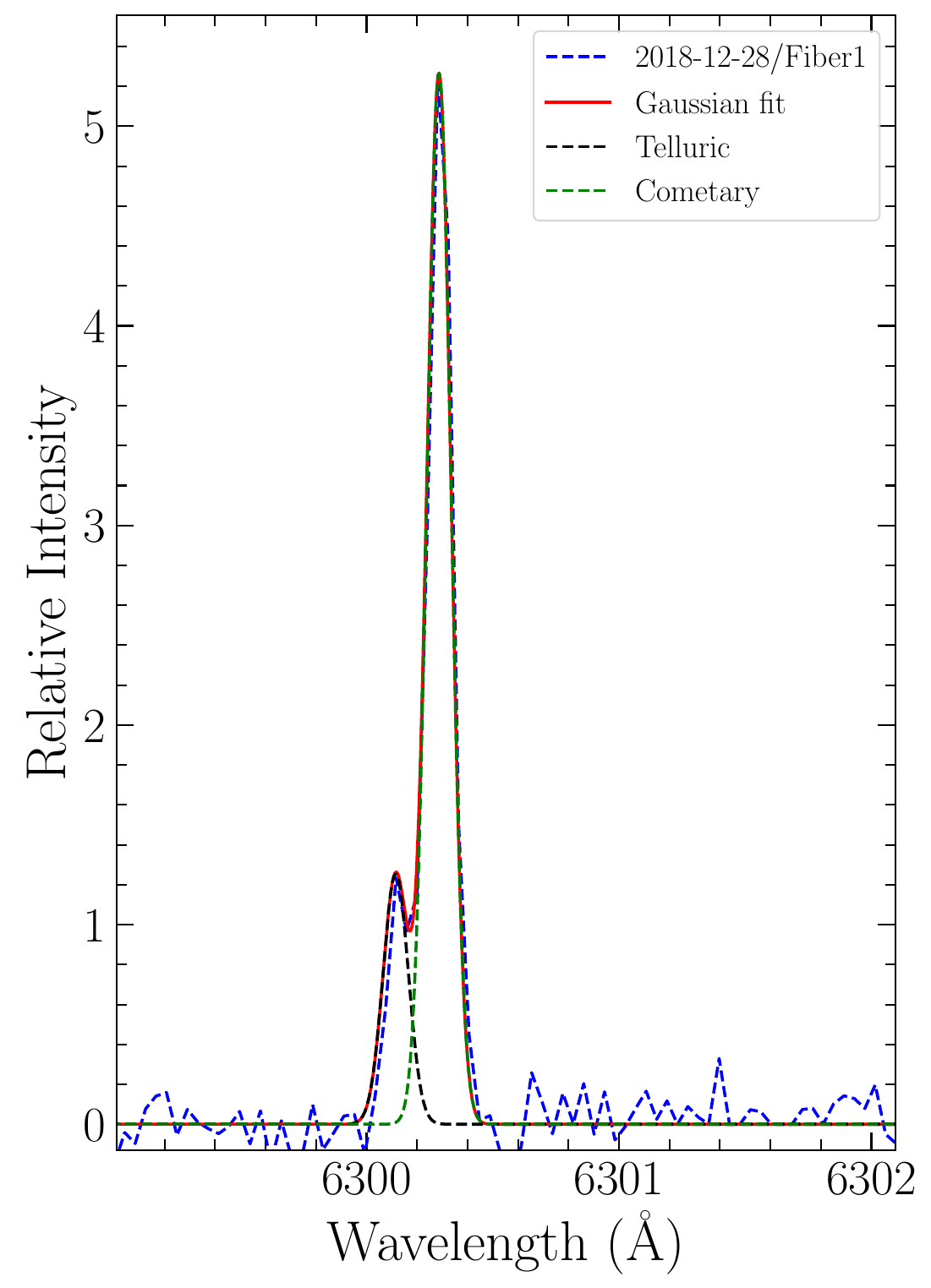}
\subcaption{{[OI] 6300 \AA }}
\label{fig:46P_O_R1}
\end{subfigure}
\begin{subfigure}{0.33\textwidth}
\centering
\includegraphics[width = 0.96\linewidth]{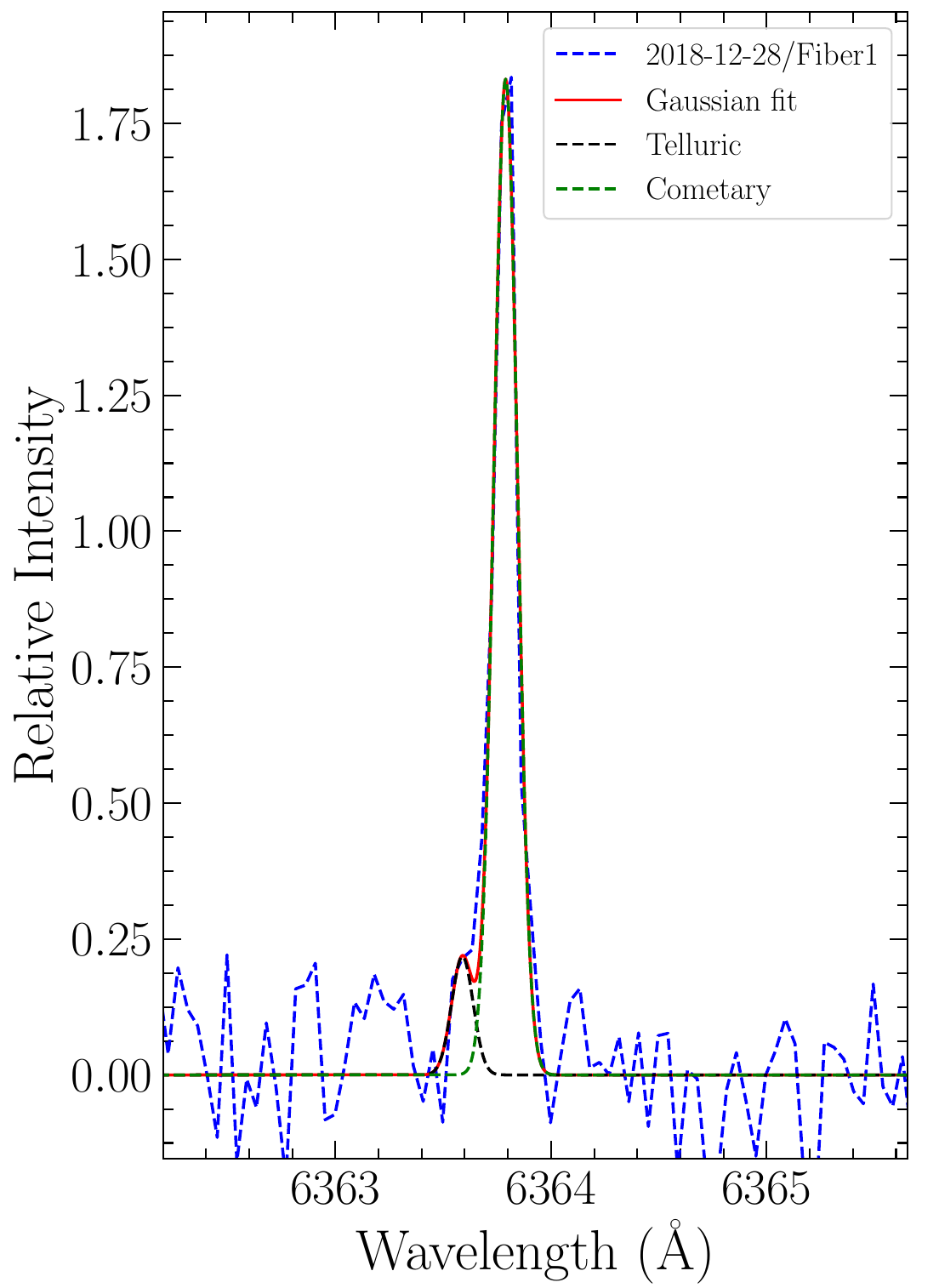}
\subcaption{{[OI]  6364 \AA}}
\label{fig:46P_O_R2}
\end{subfigure}
\caption{Green and red doublet [OI] forbidden lines observed in 46P on 2018-12-28 through the high-resolution mode.}
\label{fig:46P_O_highres}
\end{figure*}

The major source of the Oxygen lines is considered to be H$_2$O if the G/R is observed to be $\sim$ 0.09 or less and the major source is considered to be CO$_2$ if the G/R is as high as $\sim$ 0.6 \citep{Raghuram_2014, Bhardwaj_Raghuram_2012}. The major catch here is that, for comets in which Oxygen lines are detected at far-off heliocentric distances, the major source would be CO$_2$ as the temperature would not be high enough for water molecules to sublimate in a large amount \citep{Decock_2013}. As the comet moves in, water ice sublimation increases, reducing the G/R. This means that the G/R ratio also depends on the heliocentric distance. But, \cite{decock_2015} observed a nucleocentric dependence of the G/R ratio in multiple comets. The G/R ratio was seen to be high ($\sim 0.1-0.2$) close to the nucleus within 100 km and dropped drastically at about 1000 km ($\sim 0.04-0.06$).  Such differences in G/R ratio with nucleocentric distance are observed due to the high quenching of Oxygen atoms by the high density of H$_2$O molecules present in the inner coma \citep{Bhardwaj_Raghuram_2012}. Such quenching mostly affects the atoms present in the O($^1$D) level as its lifetime is much larger than those in O($^1$S) \citep{Raghuram_2014}, causing the increase in G/R close to the nucleus. At distances far enough from the nucleus where quenching is not a major factor, the actual value of the G/R ratio is observed. \cite{67P_Oxygen_GR} have also discussed the probability of including Oxygen molecules as a possible source of these Oxygen emissions. But, the effect of the Oxygen molecules would only be predominant in the near nucleus region within 400 km.

The forbidden oxygen emission lines, the Green (G) and Red doublet (R) originating from the O($^1$S) and O($^1$D) levels, respectively, are present both in comets and the Earth's atmosphere. Since both cometary and telluric Oxygen lines fall at the same wavelength, separating them in low-resolution spectroscopic observation is nearly impossible. Even though the geocentric velocity of the comet induces a Doppler shift in the cometary Oxygen line, a reasonably high resolution is required to separate it from the telluric line. In a special case, if the comet possesses large geocentric velocity, it will be possible to separate the Oxygen lines even with a resolution of 30000. Even though the three lines fall in 3 different orders of the spectrum, considering that they are continuum normalised and fall more or less on the central part of the CCD, the sensitivity effect can be considered minimal. On the other hand, the presence of certain C$_2$ lines blended along with the 5577.339 \AA~ Oxygen lines may induce minor effects on the computed G/R ratio. The synthetic spectrum of C$_2$ fabricated, as mentioned in \cite{decock_2015}, is to be removed from the observed spectrum to nullify this effect.

The effect of these blends varies from comet to comet as the Carbon chain depletion varies in each of them. In the current case, a basic ratio computation has been performed to understand the general characteristics of the composition present in the different comets. Even though a systematic study of change in G/R with nucleocentric distance is not possible in the case of HESP observations, the presence of 2 fibres 13 arcsec apart makes it feasible to measure the ratio at two different locations of the coma. This can prove important in analysing the effects of quenching on the G/R ratio close to and away from the nucleus.

In addition, using equations 7 and 9 in \cite{Decock_2013}, we have computed the intrinsic line widths and line velocity widths for the green and red doublets to compare the broadening and analyse their variation with heliocentric and nucleocentric distance. The measured respective intensities, red doublet line intensity ratios, and FWHMs in both the fibres for the sample of comets observed in this work are given in Table \ref{tab:fwhm}.

{\underline{\textbf{V2}}: }As mentioned in Table.\ref{tab:observation}, comet V2 was observed across three epochs with a resolution of 30000. Out of the 3, the comet's geocentric velocity was large enough during the first two epochs to induce a high enough Doppler shift to separate the cometary and telluric Oxygen lines. Figure.\ref{fig:C2015V2_OI_fiber2} illustrates the Oxygen green and red doublets observed in V2 on 2017-02-22. The substantial separation between the Oxygen lines makes it easier to measure the intensities corresponding to the three lines with the help of Gaussian fitting and, hence, compute the G/R ratio. Table.\ref{tab:oxygen_GR} provides the computed values for the two fibre inputs corresponding to the two epochs of observation. In the case of V2, during both epochs, the geocentric distance was fairly large, making it difficult to probe the inner coma.

Since the physical aperture of Fibre 1 itself was greater than 1000 km during both epochs, the G/R value corresponding to both Fibre 1 and 2 would, in a way, represent the actual value unaffected by the quenching effects. Hence, from the computed G/R values for the two epochs, it can be implied that the contribution of H$_2$O for producing the Oxygen lines increased drastically as the comet was approaching perihelion. The consistency in the measured values in both the fibres for the second epoch shows that the major effects of collisional quenching were well within $\sim$1000 km.

\begin{figure}
\centering
\includegraphics[width = 0.95\linewidth]{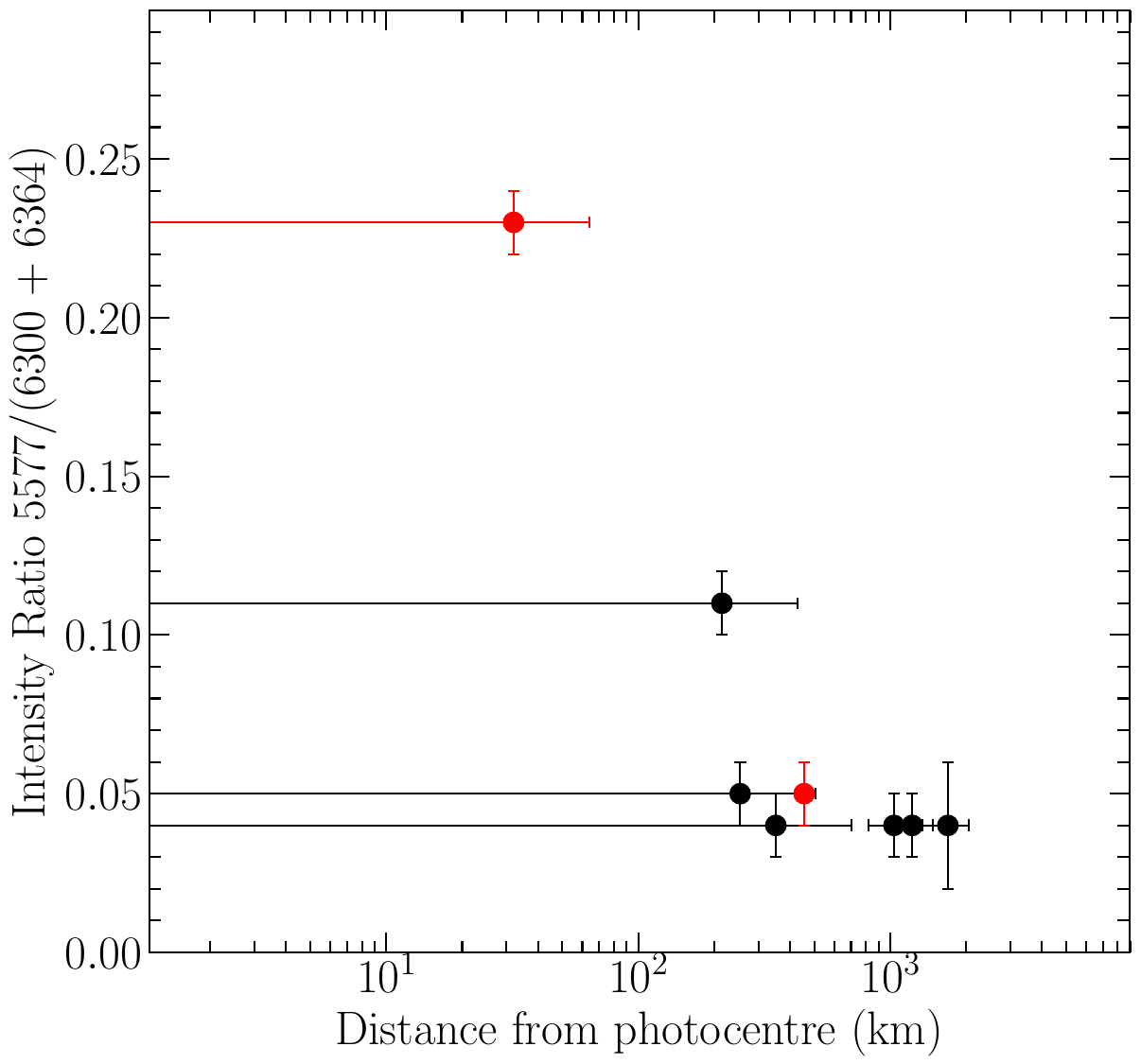}
\caption{Illustration of variation in G/R ratio computed for comet 46P in each fibre at different epochs reported in this work (black points) and \citet{46P_jehin_highres} (red points), as a function of distance from the photocentre (refer Table~\ref{tab:oxygen_GR} for respective values).}
\label{fig:G_R_nucl}
\end{figure}

{\underline{\textbf{46P}}: }The geocentric velocity of the comet was not high enough during the second epoch as the comet had a close approach to Earth at the same time. Hence, the separation in the oxygen lines induced by the Doppler shift could not be probed at the 30000 or 60000 resolution for the observation epoch close to perihelion. The observation of 46P in the 60000 resolution on 2018-12-28 proved significant since the Oxygen line peaks could be separated. As shown in Figure.\ref{fig:46P_O_highres}, Gaussian deblending was used to measure the intensities of the cometary Oxygen lines and hence measure the G/R values. During the 2018-11-28 and 2019-01-11 epochs of observation, the geocentric velocity played a major role in separating the lines in the 30000 resolution. Still, the Gaussian deblending had to be incorporated as the separation was not evident enough, as in the case of V2. The measured G/R values corresponding to both the fibres for the three epochs are given in Table.\ref{tab:oxygen_GR}.\\
The historic closest approach of 46P proved major in comparing the G/R values in the coma close to the nucleus and away from it. During the third epoch of observation (2018-12-28), Fibre 1 looked at the very inner part of the coma, around 215 km (corresponding to the size of the fibre), and Fibre 2 looked at a region more than $\sim$ 1000 km away from the nucleus. According to the study of variation in G/R values with nucleocentric distance for multiple comets as shown in \cite{Decock_2013}, these two regions correspond to different effects of quenching. Interestingly as expected, the G/R values measured for fibres probing two far-away locations in the coma vary largely. The consistency of the values measured for Fibre 1 and Fibre 2 on the other two epochs of observation (2018-11-28 and 2019-01-11) elucidate the major effect of collisional quenching in the inner parts of the coma, well within 250 km. \\
The observed value of G/R in comet 46P, in the sky Fibre, is similar to what has been reported in \cite{46P_jehin_highres}. The difference in the values observed in the central Fibre would be a direct effect of the size of the Fibre used in both observations. \cite{46P_jehin_highres} using a 0.5$^{\prime\prime}$ fibre would be looking at the central part of the coma majorly affected by quenching effects compared to the 2.7$^{\prime\prime}$ fibre used in this work. Figure \ref{fig:G_R_nucl} depicts the variation in G/R ratio computed for comet 46P in each fibre at different epochs reported in this work and \cite{46P_jehin_highres}, as a function of distance from the photocentre. Similar to the work done by \cite{Raghuram_R2_highres}, \cite{Decock_2013}, and \cite{67P_Oxygen_GR}, further modelling is to be incorporated to use this information to understand better the phenomenon of quenching and the abundance of H$_2$O, CO$_2$ and Oxygen in the comet.\\

{\underline{\textbf{41P}}: }It was seen that the signal in various molecular bands for comet 41P was too weak to be analysed in depth. However, the strong Oxygen emission along with the high geocentric velocity of the comet at the time of observation (see Table \ref{tab:observation}) favoured the study of the G/R ratio in both the fibres. During the comet's 2017 apparition, \cite{41P_trappist} and \cite{46P_21year_waterprod} reports low activity (in both gas and dust) and water production rate, respectively. Hence, as explained by \cite{Raghuram_2014}, in this case, it can be assumed that the region affected by collisional quenching would be very close to the nucleus. Considering the geocentric distance of the comet at the time of observation, the central fibre was looking at a region close to 500 km at the centre of the coma, while the second fibre was probing a region about 2357 km away from the photocentre. From the computed G/R values for both the fibres (see Table.\ref{tab:oxygen_GR}), it can be inferred that both the central and second fibre are probing a region free of collisional quenching. Based on the conclusion provided in \cite{Raghuram_2014}, these values could also be constructively used to constrain the upper limit of relative abundance in CO$_2$ present in the comet.

\begin{table}
\centering
\caption{Observational results of G/R }
\setlength\tabcolsep{3pt}
\renewcommand{\arraystretch}{1.2}
\begin{tabular}{cccccc}
\hline
    Comet & Date of & Fibre 1 & Fibre 2 & Separation of& Reference\\ 
    & observation & (error) & (error) & Fibres [km] & \\
\hline
46P & 2018-11-28 & 0.05 & 0.04 & 1225 & This work \\
          &      & (0.01)& (0.01)    &          \\ 
46P & 2018-12-08 & 0.23 & 0.05 & 467 & {\color{blue} Moulane et al}\\
    &            & (0.01)      &  (0.01) &   & {\color{blue}(2023)}\\
46P & 2018-12-28 & 0.11 & 0.04 & 1037 & This work \\
          &      & (0.01)& (0.01)    &          \\ 
 & 2019-01-19 & 0.04 & 0.04 & 1697 & This work\\
           &      & (0.01)& (0.01)    &          \\ 
&  &  &  &  &  \\
C/2015V2 & 2017-02-22 & -- & 0.10 & 16500 & This work\\
          &            &    & (0.01)    &          \\  
          & 2017-05-02 & 0.04 & 0.04 & 9334 & This work\\
           &           &  (0.01)  & (0.01)    &          \\ 
&  &  &  &  &  \\
41P & 2017-02-22 & 0.13 & 0.08 & 2357 & This work\\
          &      & (0.05)& (0.02)    &          \\ 
\hline
\end{tabular}
\label{tab:oxygen_GR}
\end{table}

\paragraph{FWHM analysis of the [OI] lines\\}
\cite{Raghuram_2014} states that the O($^1$S) transition responsible for the green line is majorly controlled by CO$_2$ production, and the  O($^1$D) transition responsible for the red doublets are mostly governed by the H$_2$O production. The mean excess energy released in photodissociation of CO$_2$ is higher than in the case of H$_2$O \citep{Raghuram_Bhardwaj_2013}.  Hence, the line width velocities of the green oxygen lines measured at the photocentre are often reported to be higher than the values estimated from the red doublets with similar intrinsic line widths \citep{Decock_2013, Cochran_2008}. 

In this work, similar to the reports in other literature \citep{Cochran_2008, Decock_2013}, the measured line widths at the photocentre for all comets point to a wider width for the green line (about 1 km s$^{-1}$ higher) in comparison to the red doublets(see table \ref{tab:fwhm}). Figure \ref{fig:FWHM} illustrates the similarity of line width velocity observed for the red doublets (top panel) and the comparison of the line width velocity of the green line with the average width of red lines (bottom panel).

\cite{decock_2015} states that the line width velocity of the green line drops significantly with nucleocentric distance. 
Similarly, in this work, the line width velocity of the green line measured in fibres 1 and 2 drops and becomes comparable to that of the red doublets. 
This is clearly seen in the bottom panel of Figure \ref{fig:FWHM}, where the open circles depicting measurements from fibre 2 show near equivalent line widths for the green and red doublets except for the fibre 2 measurement of V2 on 2017-02-22. This observed higher width of green line in fibre 2 for V2 would imply a larger contribution from CO$_2$ than from H$_2$O in comparison to the other epochs where the comet came closer to the Sun.

The measurements for comet 46P in both fibres for 2018-12-28 show a very high value for the line widths compared to the other comets. Considering that 46P was reported to be a hyperactive comet \citep{46P_jehin_highres, 46P_knight_schleicher, 46P_nowater_icegrains}, the higher width of the lines could be a direct effect of the higher water production activity \citep{decock_2015}. Keeping aside the 46P 2018-12-28 observation, the measured average of the velocity widths at the photocentre for all comets in this work is 2.43 $\pm$ 0.28 for the 5577 \AA, 1.70 $\pm$ 0.33 for the 6300 \AA~ line and 1.66 $\pm$ 0.38 for the 6364 \AA~ line. Furthermore, the red doublet intensity ratio is comparable for all the comets, with an average of 3.05 $\pm$ 0.13 for the sample reported in this work. The above-mentioned findings are in good agreement with the corresponding values reported in \citep{Cochran_2008} for a sample of 8 comets and in \cite{Decock_2013} for a sample of 12 comets.

\begin{figure}
\centering
\includegraphics[width = 0.95\linewidth]{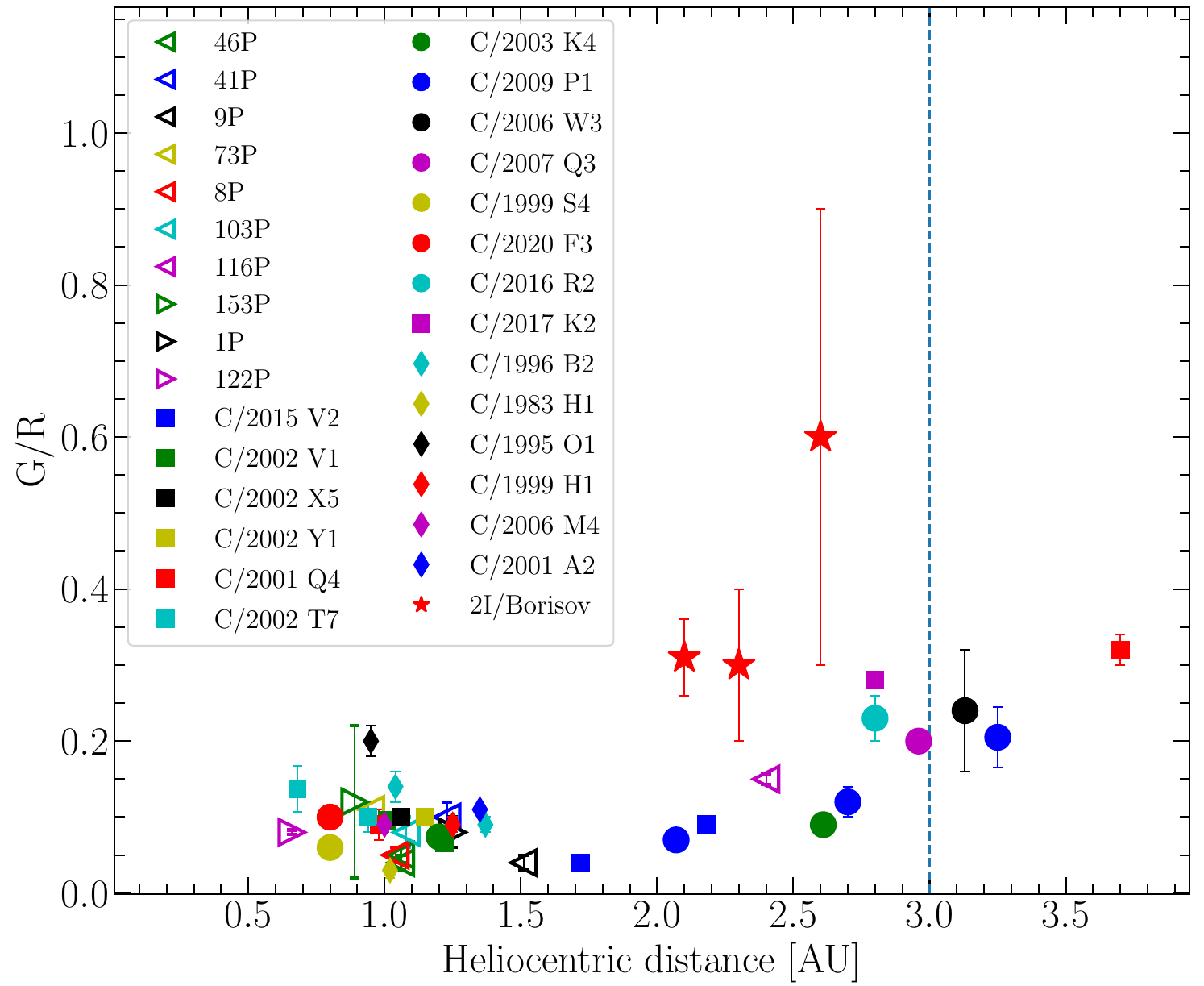}
\caption{Ratio between the green and the sum of the two red forbidden
oxygen lines (G/R) in comets 46P, 41P and C/2015 V2 (reported in this work), compared to that measured for a sample of Solar System comets reported in \citet{Decock_2013}, \citet{Cochran_2008}, \citet{G_R_116P}, \citet{G_R_153P}, \citet{Zhang_2001}, \citet{G_R_halley}, \citet{G_R_hyakutake}, \citet{G_R_iras_araki}, \citet{G_R_116P}, \citet{G_R_1999S4}, \citet{G_R_2006W3_2007Q3}, \citet{G_R_K2}, \citet{Opitom_R2_highres}, \citet{F3_highres} 
and the first interstellar comet 2I/Borisov reported in \citet{opitom_borisov}. Open markers represent short-
period comets and solid markers represent long-period comets. The vertical dashed line at 3 AU represents the distance out to which the sublimation of water is strongly decreasing \citep{crovisier}.
}
\label{fig:G_R_rh}
\end{figure}

The G/R values of the comets reported in this work, along with those reported for other comets in the literature (see Table \ref{tab:gbyr}), have been illustrated in Figure \ref{fig:G_R_rh} as a function of the heliocentric distance at the time of observation. Excluding the values reported for the interstellar comet 2I/Borisov, it is clear that the G/R value reduces as the comet comes closer to the Sun owing to the increased production of H$_2$O.  Observations of the ratio for a single comet over a large range of heliocentric distances would be necessary to identify any systematic dependence on heliocentric distance. 

\section{Conclusion}
While only a small fraction of the known comets have been observed in the spectroscopic technique, only a few among them have been observed with very high resolution to probe the different vibrational/rotational lines present in the different molecular emission bands or to compute the G/R ratio and OPR. In this work, we have discussed the high-resolution spectroscopy of 4 comets originating from different reservoirs. The specific conclusions of the work are given below:
\begin{enumerate}
    \item Four comets have been observed and analysed using both modes of the high-resolution Hanle Echelle SPectrograph (HESP) on the 2 m Himalayan Chandra Telescope, Hanle.  These are the first comet observations using the HESP instrument. 
    \item For the observed comets, we have identified the lines present in the various molecular bands during different epochs.
    \item The variation in the strength of the emission lines present in the different major molecular bands along the orbit of these comets has been visualised.
    \item The green and red doublet forbidden oxygen lines in comets C/2015 V2, 46P, and 41P could be resolved from the telluric emissions and have been analysed to compute their intensities, intrinsic line velocity and G/R ratio.
    \item The ratio of the red doublet intensity ({I$_{6300}$/I$_{6364}$}) measured to be 3.05 $\pm$ 0.13, along with a wider line width of the green line (for measurements at the photocentre) with average 2.43 $\pm$ 0.28 km s$^{-1}$ compared to the red doublet lines, 6300 \AA~and 6364 \AA, with average 1.70 $\pm$ 0.33 km s$^{-1}$ and 1.66 $\pm$ 0.38 km s$^{-1}$ respectively, are seen to be in agreement to the corresponding values reported in \cite{Cochran_2008} and \cite{Decock_2013}.
    \item The G/R ratio computed for all three comets point to H$_2$O being a major source of Oxygen emissions. The high G/R values obtained for comets C/2015 V2 and 41P outside the region affected by quenching could be used to constrain the upper limit of relative abundance of CO$_2$ present in them.
    \item Comet 46P, having had a historic close approach to Earth, enabled the analysis of the G/R ratio at two locations in the coma - one at the photocenter and the other about 1000 km away. Detailed analysis and visualisation of the variation in G/R with nucleocentric distance indicates the strong effects of quenching in the inner coma. Further modelling, which is beyond the scope of this work, is necessary to understand better the phenomenon of quenching and the abundance of various molecules like CO$_2$ and H$_2$O.
    \item The high SNR in 46P high-resolution spectra helped in probing the different NH$_2$ bands present. They were used to compute the OPR to have a general understanding of the spin temperature present in the comet. 
    \item Detailed modelling techniques are to be employed with NH$_2$ and Oxygen line analysis to remove all the possible blends to obtain the accurate line ratios.  
\end{enumerate}

\section*{Acknowledgements}
We thank the referee, Dr Pamela Cambianica, for the valuable comments and suggestions that have significantly improved the quality of the manuscript. Work at PRL is supported by the Department of Space, Govt. of India. AB was JC Bose Fellow during the period of this work.\\
We thank the staff of the Indian Astronomical Observatory, Hanle and the Centre For Research \& Education in Science \& Technology, Hoskote, who made these observations possible. The facilities at IAO and CREST are operated by the Indian Institute of Astrophysics, Bangalore.  

\section*{Data Availability}

The data underlying this article will be shared on reasonable request to the corresponding author. 

\bibliography{reference.bib}{}
\bibliographystyle{mnras}

\appendix
\section{The measured intrinsic FWHM and red doublet ratios}

\begin{figure}
\centering
\includegraphics[width = 0.93\linewidth]{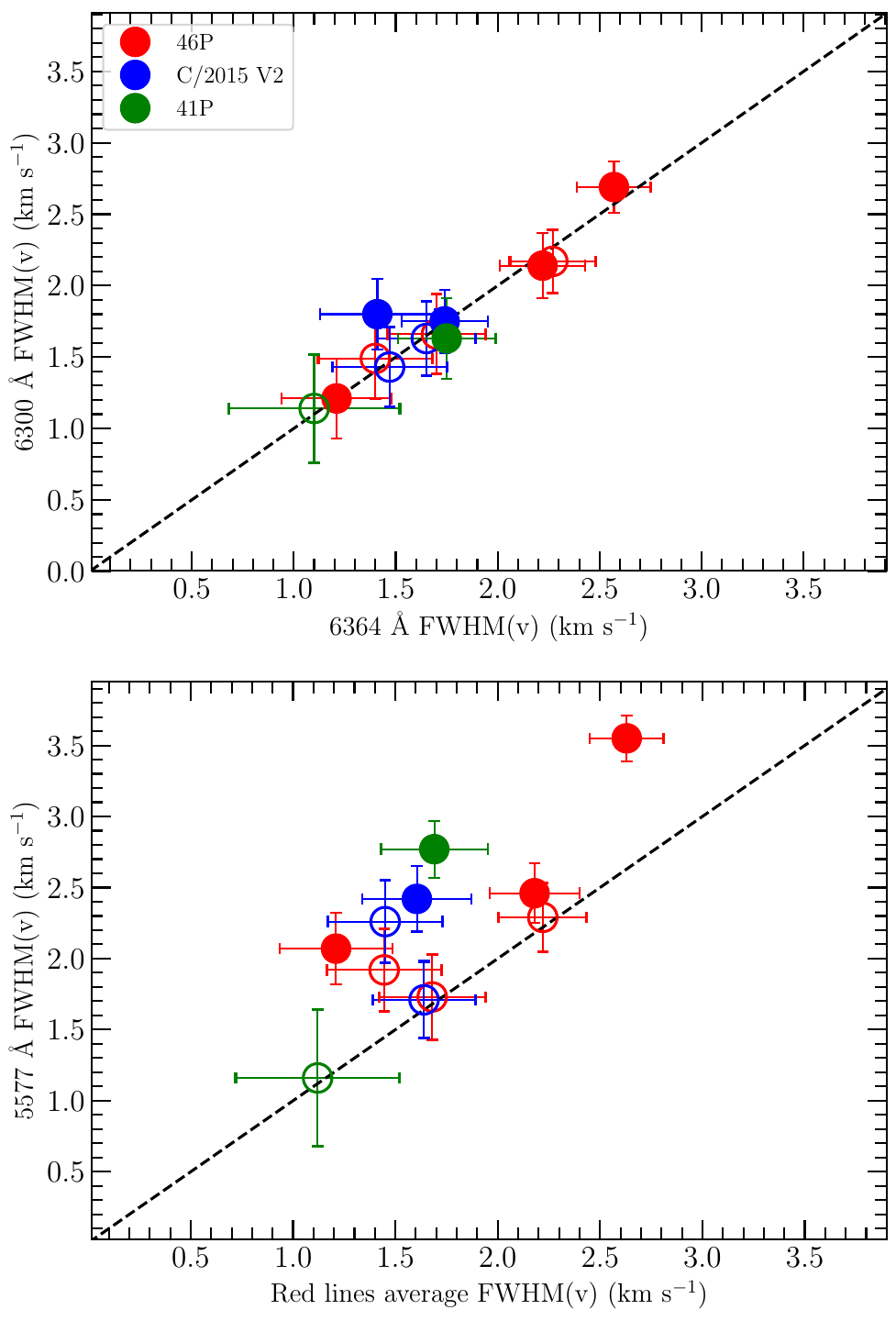}
\caption{\textit{Top panel:} The correlation of the line width velocities measured for the 6300 \AA~ and 6364 \AA~ Oxygen lines; \textit{Bottom panel:} The correlation of the line width velocities measured for the 5577 \AA~ and average of the red doublet Oxygen lines. Different comets are marked by different colours, while the solid circles depict measurements from fibre 1, and open circles are for measurements from fibre 2.}
\label{fig:FWHM}
\end{figure}

The intrinsic line width velocity of the forbidden oxygen green and red doublet lines corresponding to both the fibres were measured for all the comets to facilitate the better understanding of the origin of these lines. The relative intensities of the individual lines, the ratio of the intensities for the red doublets and the measured intrinsic line width velocities are given in Table \ref{tab:fwhm}. 
Line width measurements at the photocenter for all comets indicate a broader width for the green line, approximately 1 km s$^{-1}$ higher than the red doublets with similar line widths. Top panel of Figure \ref{fig:FWHM} visually demonstrates the comparable line width velocities of the red doublets and the bottom panel highlights the wider nature of the green line when compared to the average width of the red lines. Our study reveals a drop in the line width velocities of the green line between fibres 1 and 2, aligning the latter values with the widths of the red doublets. This is evident in the bottom panel of Figure \ref{fig:FWHM}, where measurements from fibre 2 display nearly identical line widths for the green and red doublets, except for the fibre 2 measurement of V2 on 2017-02-22. The increased width of the green line in fibre 2 for V2 suggests a greater contribution from CO$_2$ compared to H$_2$O during this epoch when the comet was farther away from the Sun compared to the other epoch.

\begin{table*}
\centering
\caption{Measured intensity ratio of the red doublets and intrinsic velocity widths (km s$^{-1}$) for the three forbidden oxygen lines in the spectra of each comet.}
\setlength\tabcolsep{3pt}
\renewcommand{\arraystretch}{1.2}
\begin{tabular}{|l|c|c|c|c|c|c|c|c|c|c|c|}
\hline
  \multicolumn{1}{|c|}{Comet name} &
  \multicolumn{1}{c|}{Date} & \multicolumn{1}{c|}{r$_h$} & \multicolumn{1}{c|}{$\Delta$} &
  \multicolumn{1}{c|}{fibre} & \multicolumn{3}{c|}{Intensity} &
  \multicolumn{1}{c|}{$\frac{\mathrm{I}_{6300}}{\mathrm{I}_{6364}}$} &
  \multicolumn{3}{c|}{FWHM$_\textit{intrinsic}$ (km s$^{-1}$)} \\
   & \multicolumn{1}{|c|}{[UT]} & \multicolumn{1}{|c|}{[AU]} & \multicolumn{1}{|c|}{[AU]} &  & 5577.339 \AA & 6300.304 \AA & 6363.776 \AA &  & \multicolumn{1}{|c|}{5577.339} & \multicolumn{1}{|c|}{6300.304} & \multicolumn{1}{|c|}{6363.776}\\
   \hline
  46P & 2018-11-28 & 1.07 & 0.13 & 1 & 0.28 & 4.45 & 1.40 & 3.16 & 2.46$\pm$0.21 & 2.14$\pm$0.23 & 2.22$\pm$0.21\\
   &  & & & 2 & 0.23 & 4.32 & 1.46 & 2.95 & 1.73$\pm$0.30 & 1.66$\pm$0.28 & 1.70$\pm$0.24\\
   & 2018-12-28 & 1.08 & 0.11 & 1 & 0.23 & 1.60 & 0.55 & 2.91 & 3.55$\pm$0.16 & 2.69$\pm$0.18 & 2.57$\pm$0.18\\
   &   & & & 2 & 0.04 & 0.68 & 0.23 & 2.96 & 2.29$\pm$0.24 & 2.17$\pm$0.22 & 2.27$\pm$0.21\\
   & 2019-01-11 & 1.13 & 0.18 & 1 & 0.06 & 1.10 & 0.35 & 3.14 & 2.07$\pm$0.25 & 1.21$\pm$0.28 & 1.21$\pm$0.27\\
   &  & & & 2 & 0.02 & 0.34 & 0.11 & 3.09 & 1.92$\pm$0.29 & 1.49$\pm$0.28 & 1.40$\pm$0.28\\
   &  & & &  & & & & &  &  & \\
  C/2015 V2 & 2017-02-22 & 2.18 & 1.78 & 1 & - & 0.36 & 0.12 & 3.00 & - & 1.75$\pm$0.22 & 1.74$\pm$0.21\\
   &  & & & 2 & 0.12 & 0.92 & 0.30 & 3.07 & 2.26$\pm$0.29 & 1.43$\pm$0.28 & 1.47$\pm$0.28\\
   & 2017-05-02 & 1.72 & 0.99 & 1 & 0.14 & 2.56 & 0.87 & 2.95 & 2.42$\pm$0.23 & 1.80$\pm$0.25 & 1.41$\pm$0.28\\
   &  & & & 2 & 0.08 & 1.44 & 0.46 & 3.13 & 1.71$\pm$0.27 & 1.63$\pm$0.26 & 1.65$\pm$0.24\\
   &  & & &  &  & & & &  &  & \\
  41P & 2017-02-22 & 1.23 & 0.25 & 1 & 0.08 & 0.47 & 0.14 & 3.35 & 2.77$\pm$0.20 & 1.63$\pm$0.28 & 1.75$\pm$0.24\\
   &  & & & 2 & 0.18 & 1.62 & 0.56 & 2.89 & 1.16$\pm$0.48 & 1.14$\pm$0.38 & 1.10$\pm$0.42\\
\hline\end{tabular}
\label{tab:fwhm}
\end{table*}

\begin{table*}
\centering
\caption{Comets with measured G/R and the corresponding heliocentric distance. The Table includes measurements from the literature and the values reported in this work.}
\setlength\tabcolsep{3pt}
\renewcommand{\arraystretch}{1.2}
\begin{tabular}{lcll|lcll}
\cmidrule(r){1-4}
\cmidrule{5-8}
\multicolumn{1}{l}{Comet}&\multicolumn{1}{c}{r$_h$}&\multicolumn{1}{l}{G/R}&\multicolumn{1}{l|}{Reference}&\multicolumn{1}{|l}{Comet}&\multicolumn{1}{c}{r$_h$}&\multicolumn{1}{l}{G/R}&\multicolumn{1}{l}{Reference}\\
\cmidrule(r){1-4}
\cmidrule{5-8}

46P/Wirtanen&1.06&0.04 $\pm$ 0.01& This work&C/2009 P1 (Garradd)&3.25&0.205 $\pm$ 0.04 &\cite{Decock_2013}\\
46P/Wirtanen&1.06&0.05 $\pm$ 0.01& \cite{46P_jehin_highres}&C/2009 P1 (Garradd)&2.70&0.12 $\pm$ 0.02 &\cite{Decock_2013}\\
41P/T–G–K&1.23&0.10 $\pm$ 0.02&This work&C/2009 P1 (Garradd)&2.07&0.07 $\pm$ 0.003 &\cite{Decock_2013}\\
116P&2.40&0.15 $\pm$ 0.007&\cite{G_R_116P}&C/2002 X5 (Kudo-Fujikawa)&1.06&0.1 $\pm$0.003 &\cite{Decock_2013}\\
9P/Tempel 1&1.51&0.04 $\pm$ 0.01&\cite{Decock_2013}&C/2002 Y1 (Juels-Holvorcem)&1.15&0.10 $\pm$0.01 &\cite{Decock_2013}\\
73P-C/S-W 3&0.95&0.11&\cite{Decock_2013}&C/2007 Q3 (Siding Spring)&2.96&0.20&\cite{G_R_2006W3_2007Q3}\\
8P/Tuttle&1.04&0.05 $\pm$ 0.01&\cite{Decock_2013}&C/2006 W3 (Christensen) &3.13&0.24 $\pm$ 0.08 &\cite{G_R_2006W3_2007Q3}\\
103P/Hartley 2&1.08&0.08 $\pm$ 0.02&\cite{Decock_2013}&C/1999 S4 (LINEAR)&0.80&0.06 $\pm$ 0.01&\cite{G_R_1999S4}\\
153P/Ikeya–Zhang&0.89&0.12 $\pm$ 0.1&\cite{G_R_153P}&C/2020 F3 (NEOWISE)&0.80&0.10 $\pm$ 0.01&\cite{F3_highres}\\
1P/Halley&1.25&0.08 $\pm$ 0.02&\cite{G_R_halley}&C/2017 K2 (PanSTARRS)&2.80&0.28 $\pm$ 0.01&\cite{G_R_K2}\\
122P/de Vico&0.66&0.08 $\pm$ 0.003 &\cite{Cochran_2008}&C/1996 B2 (Hyakutake)&1.04&0.14 $\pm$ 0.02&\cite{G_R_hyakutake}\\
C/2015 V2 (Johnson)&2.18&0.09 $\pm$ 0.01&This work&C/1996 B2 (Hyakutake)&1.37&0.09 $\pm$ 0.01&\cite{Cochran_2008}\\
C/2015 V2 (Johnson)&1.72&0.04 $\pm$ 0.01&This work&C/1999 H1 (Lee)&1.25&0.08 $\pm$ 0.01&\cite{Cochran_2008}\\
C/2002 V1 (NEAT)&1.22&0.066 $\pm$ 0.01& \cite{Decock_2013}&C/2006 M4 (Swan)&1.00&0.09 $\pm$ 0.01&\cite{Cochran_2008}\\
C/2002 V1 (NEAT)&1.01&0.096 $\pm$ 0.01&\cite{Decock_2013}&C/2001 A2 (LINEAR)&1.35&0.11&\cite{Cochran_2008}\\
C/2002 V1 (NEAT)&0.98&0.09&\cite{Cochran_2008}&C/1983 H1 (IRAS Araki)&1.02&0.03 $\pm$ 0.01&\cite{G_R_iras_araki}\\
C/2001 Q4 (NEAT)&0.98&0.09 $\pm$ 0.02 &\cite{Cochran_2008}&C/1995 O1 (Hale-Bopp)&0.95&0.2 $\pm$ 0.02 &\cite{Zhang_2001}\\
C/2001 Q4 (NEAT)&3.70&0.32 $\pm$ 0.02 &\cite{Decock_2013}&C/2016 R2 (PanSTARRS)&2.80&0.23 $\pm$ 0.03&\cite{Opitom_R2_highres}\\
C/2002 T7 (LINEAR)&0.68&0.137 $\pm$ 0.03&\cite{Decock_2013}&2I/Borisov&2.10&0.31 $\pm$ 0.05&\cite{opitom_borisov}\\
C/2002 T7 (LINEAR)&0.94&0.1 $\pm$ 0.02 &\cite{Decock_2013}&2I/Borisov&2.30&0.30 $\pm$ 0.10 &\cite{opitom_borisov}\\
C/2003 K4 (LINEAR)&2.61&0.09 $\pm$ 0.01 &\cite{Decock_2013}&2I/Borisov&2.60&0.60 $\pm$ 0.30 &\cite{opitom_borisov}\\
C/2003 K4 (LINEAR)&1.20&0.074 $\pm$ 0.01 &\cite{Decock_2013}&\\
\cmidrule(r){1-4}
\cmidrule{5-8}
\end{tabular}
\label{tab:gbyr}
\end{table*}

\end{document}